\documentclass[aps,prd,twocolumn,preprintnumbers,superscriptaddress,dblfloatfix,nofootinbib]{revtex4-1}
\usepackage[utf8]{inputenc}
\usepackage{lmodern}
\usepackage{amsmath,amssymb,mhchem}
\usepackage{graphicx}
\usepackage{url}
\usepackage{color}
\usepackage{enumitem}
\usepackage{subfigure}
\usepackage[dvipsnames]{xcolor}
\usepackage[colorlinks=true,breaklinks=true]{hyperref}
\hypersetup{allcolors=[rgb]{0.0 0.0 0.6},linkcolor=[rgb]{0.75 0.05 0.05}}
\usepackage{bm}
\usepackage{epsfig}
\usepackage{amsmath}
\usepackage{amssymb}
\usepackage{slashed}
\usepackage{color}
\usepackage{accents}
\usepackage{url}
\usepackage[dvipsnames]{xcolor}
\usepackage{ulem}

\newcommand{\exclude}[1]{}

\begin{document}

\preprint{IPMU21-0028}
\preprint{YITP-21-44} 

\title{Exploring Evaporating Primordial Black Holes with Gravitational Waves}

\author{Guillem Dom\`enech} \email{{domenech}@{pd.infn.it}}
\affiliation{INFN Sezione di Padova, I-35131 Padova, Italy} 

\author{Volodymyr Takhistov} \email{volodymyr.takhistov@ipmu.jp}
\affiliation{Kavli Institute for the Physics and Mathematics of the Universe (WPI),University of Tokyo, Chiba 277-8583, Japan}

\author{Misao Sasaki} \email{{misao.sasaki}@{ipmu.jp}}
\affiliation{Kavli Institute for the Physics and Mathematics of the Universe (WPI),University of Tokyo, Chiba 277-8583, Japan}
\affiliation{Center for Gravitational Physics, Yukawa Institute for Theoretical Physics, Kyoto University, Kyoto 606-8502, Japan} 
\affiliation{Leung Center for Cosmology and Particle Astrophysics, National Taiwan University, Taipei 10617, Taiwan}

\begin{abstract}
Primordial black holes (PBHs) from the early Universe have been connected with the nature of dark matter and can significantly affect cosmological history. We show that coincidence dark radiation and density fluctuation gravitational wave signatures associated with evaporation of $\lesssim~10^9$~g PBHs can be used to explore and obtain important hints about the formation mechanisms of spinning and non-spinning PBHs spanning orders of magnitude in mass-range, which is challenging to do otherwise.
\end{abstract}

\date{\today}

\maketitle

Primordial black holes (PBHs) that could have formed in the early Universe prior to galaxies and stars can constitute a significant fraction of the dark matter (DM), can significantly affect cosmological history and have been associated with a variety of observational signatures~(e.g.~\cite{Zeldovich:1967,Hawking:1971ei,Carr:1974nx,Chapline:1975ojl,Meszaros:1975ef,Carr:1975qj,GarciaBellido:1996qt,Kawasaki:1997ju,Khlopov:2008qy,Frampton:2010sw,Bird:2016dcv,Kawasaki:2016pql,Carr:2016drx,Inomata:2016rbd,Pi:2017gih,Inomata:2017okj,Garcia-Bellido:2017aan,Inoue:2017csr,Georg:2017mqk,Inomata:2017bwi,Kocsis:2017yty,Ando:2017veq,Niikura:2017zjd,Cotner:2016cvr,Cotner:2019ykd,Cotner:2018vug,Sasaki:2018dmp,Carr:2018rid,Flores:2020drq,Deng:2017uwc,Kusenko:2020pcg,Lu:2020bmd,Fuller:2017uyd,Kusenko:2019kcu,Takhistov:2020vxs,Kawasaki:2018daf,Takhistov:2017bpt,Takhistov:2017nmt,Laha:2020vhg,Sugiyama:2020roc,Takhistov:2021aqx}). Depending on formation scenario, PBHs can span many orders of magnitude in mass.

Very light PBHs with masses $\lesssim 10^{-15}$~g evaporate through Hawking radiation~\cite{Hawking:1974sw} on time scales shorter than the age of the Universe and thus do not contribute themselves directly to the DM abundance~\cite{Page:1976df,Page:1977um}. 
Evaporating PBHs can copiously emit constituents from the whole particle spectrum, including production of particles from the Standard Model extensions and the dark sector. Products of evaporating PBHs have been associated with the observed matter-antimatter asymmetry, DM and dark radiation (DR)~\cite{Fujita:2014hha,Lennon:2017tqq,Morrison:2018xla,Masina:2020xhk,Masina:2020xhk,Hooper:2019gtx,Hooper:2020evu,Lunardini:2019zob,Gondolo:2020uqv,Khlopov:2008qy}. PBHs spanning range $\sim 10^{-5} - 10^{9}$~g, from the Planck mass to the mass relevant for Big Bang Nucleosynthesis (BBN), remain unconstrained~(see e.g.~\cite{Carr:2020gox}).

PBHs have been often considered as non-rotating (Schwarzschild)~\cite{Chiba:2017rvs,DeLuca:2019buf,Mirbabayi:2019uph}. However, PBHs can be formed with significant spin (Kerr BHs)~\cite{Amendola:2017xhl,Flores:2020drq,Domenech:2021uyx, Harada:2016mhb,Kokubu:2018fxy,Cotner:2016cvr,Cotner:2019ykd,Cotner:2018vug} as well as acquire spin via accretion~\cite{DeLuca:2020bjf} or hierarchical mergers~\cite{Fishbach:2017dwv}. 
Aside mass and charge, spin constitutes a fundamental conserved BH parameter.
BH spin affects Hawking radiation as well as the BH lifetime, with potential consequences for observations~\cite{Arbey:2019vqx,Dong:2015yjs,Kuhnel:2019zbc,Arbey:2019jmj,Bai:2019zcd,Dasgupta:2019cae,Laha:2020vhg}.
BH rotation generally increases the emission and favors particle production with larger spin~\cite{Hawking:1974sw,Page:1976ki,Page:1977um,Taylor:1998dk}. As shown in Ref.~\cite{Hooper:2020evu} and confirmed by detailed numerical analyses~\cite{Arbey:2021ysg,Masina:2021zpu}, spin-2 graviton emission is significantly enhanced in rotating evaporating BHs, which dominates the resulting copious production of DR when PBHs of $\lesssim 10^9$~g mass dominate the energy density and that will affect BBN and cosmic microwave background radiation (CMB) observations. Note that the emitted gravitons correspond to extremely high frequency gravitational waves (GWs), which are undetectable with current or near-future interferometer experiments~(e.g.~\cite{Dong:2015yjs,BisnovatyiKogan:2004bk}).

In this work we show that coincidence observations of DR, associated primarily with PBH Hawking radiation graviton production, as well as secondary GWs from PBH density fluctuations allow us to explore and distinguish formation mechanisms of spinning and non-spinning evaporating PBHs, opening an unprecedented window into the unconstrained $\lesssim 10^9$~g PBH mass range.

On small scales, the inhomogeneous distribution of PBHs leads to density fluctuations \cite{Papanikolaou:2020qtd}. As light PBHs dominate the Universe, the initial isocurvature fluctuations are converted into curvature perturbations. During the era of PBH domination, density fluctuations grow until PBHs evaporate completely. The PBH evaporation process, which is effectively instantaneous (i.e.~sudden reheating approximation), transforms large density fluctuations into radiation and yields large pressure waves \cite{Inomata:2019ivs}. This induces a significant quadrupole moment and, therefore, results in significant GW production~\cite{Domenech:2020ssp} (see also~\cite{Inomata:2020lmk}).

\begin{figure*}[ht]
 \centering
 \includegraphics[trim={0mm 0mm 0mm 0},clip,width=0.485\textwidth]{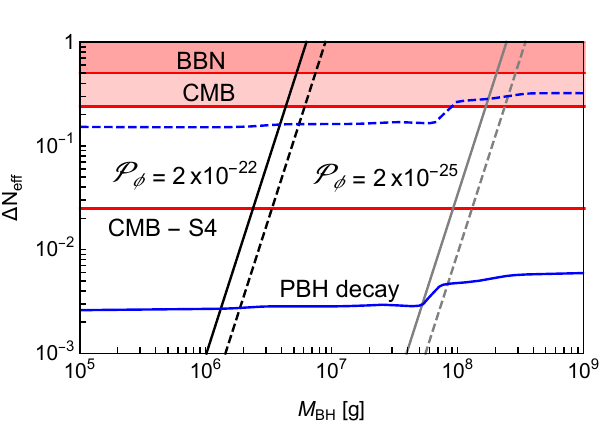} \quad
 \includegraphics[trim={0mm 0mm 0mm 2mm},clip,width=0.43\textwidth]{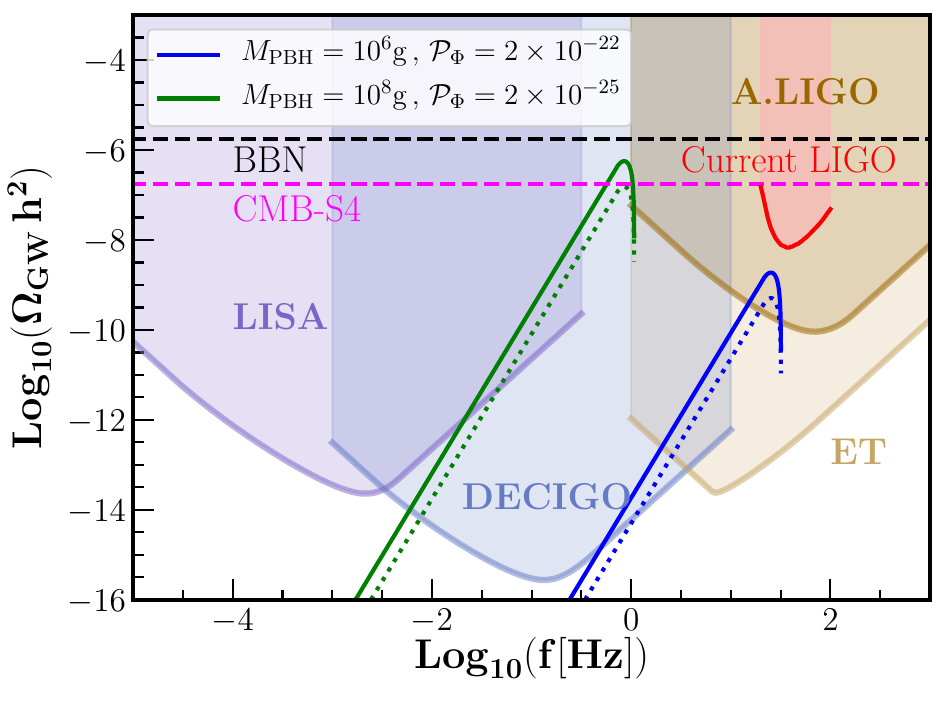}
 \caption{[Left] Contributions to $\Delta N_{\rm eff}$ from Hawking evaporation production (``PBH decay'', blue) taken from Ref.~\cite{Arbey:2021ysg}, as well as induced GWs (${\cal P}_\Phi = 2\times 10^{-22}$ in black, ${\cal P}_\Phi = 2\times 10^{-25}$ in gray) for Schwarzschild ($a_{\ast} = 0$, solid line) and Kerr ($a_{\ast} = 0.9999$, dashed).
  The 95\% C.L. limits on $\Delta N_{\rm eff}$ from CMB (shaded red) are taken from~\cite{Aghanim:2018eyx} and
the 95\% limit on BBN from~\cite{Arbey:2021ysg}, the prospective CMB-S4 sensitivity is shown with horizontal red line~\cite{Abazajian:2016yjj}.
 [Right]  GW spectral density induced by early isocurvature fluctuations in the PBH reheating scenario in terms of frequency. The blue and green solid lines correspond to ($M_{\rm BH} = 10^6$~g, ${\cal P}_\Phi = 2\times 10^{-22}$) and ($M_{\rm BH} = 10^8$~g, ${\cal P}_\Phi = 2\times 10^{-25}$), respectively. GWs from Kerr PBHs are shown with dashed lines. Note that for this choice of parameters we respectively have $k_{\rm eq}/k_{\rm reh}\sim270,1000$ and, therefore, the Universe reaches complete PBH domination. We also display the power-law integrated sensitivity curves for LISA, DECIGO, ET and Advanced LIGO experiments~\cite{Thrane:2013oya}. In red we show the latest upper bounds on the stochastic GW background from the LIGO/VIRGO/KAGRA collaboration \cite{Abbott:2021xxi}.}
 \label{fig:neffGW}
\end{figure*}

We now discuss stochastic background GW production from PBH evaporation, including effects of spin, in generality as applicable to a large class of models.
Consider a generic (dimensionless) power spectrum in wavenumber $k$ of fluctuations in the gravitational potential $\Phi$ during a matter-dominated stage described by
\begin{align}
{\cal P}_\Phi(k)={\cal A}_\Phi \left(\frac{k}{k_{\rm UV}}\right)^{-n} \Theta(k_{\rm UV}-k)~,
\end{align}
where $A_\Phi$ is the amplitude, $n$ is the spectral index and $k_{\rm UV}$ is an ultraviolet cut-off. This type of spectrum includes the typical nearly scale-invariant spectrum $(n = 0)$ found in models of inflation. As we show later, $n \sim 1$ corresponds to PBH density fluctuations.

In generality, induced GWs are generated at second order in cosmological perturbation theory \cite{tomita,Matarrese:1992rp,Matarrese:1993zf,Ananda:2006af,Baumann:2007zm,Saito:2009jt}. In terms of Feynman diagrams, one can envision graviton propagator with a scalar loop inside. Given an initial spectrum of fluctuations one can then find the induced GW spectrum by multiplying with the appropriate kernel and integrating over the internal momenta.
In the approximation of nearly sudden reheating, the contribution to the induced GWs at the peak can be schematically estimated as
\begin{align}\label{eq:omegaGW}
\Omega_{\rm GW}(k&=k_{\rm UV})\sim \int d\left(\tfrac{q}{k_{\rm UV}}\right){\cal P}_\Phi^2\left({q}\right) \overline{I^2\left(\tfrac{q}{k_{\rm UV}},\tfrac{q}{k_{\rm reh}}\right)}\nonumber\\&\sim {\cal P}_\Phi^2(k=k_{\rm UV}) S_\Phi^4({k}/{k_{\rm reh}})\left(\frac{k_{\rm UV}}{k_{\rm reh}}\right)^7\,,
\end{align}
where $q$ is the internal scalar loop momentum and $S_\Phi(k/k_{\rm reh})$ is a suppression factor to account for the finite width of the transition, so that $S_\Phi\to 1$ in the instantaneous transition limit. We refer for computation details to Ref.~\cite{Inomata:2019ivs,Domenech:2020ssp}. In the last step, we first used the fact that in the sudden reheating limit (associated with $k_{\rm reh}$), the conformal time derivative of $\Phi$, $\Phi'$, jumps from $\Phi'=0$ in a matter-dominated stage to $\Phi'\neq 0$ in radiation-dominated stage. This yields a significant amplitude of $\Phi'$ right after reheating and the kernel is essentially given by $I\sim \left({\Phi'}/{{\cal H}}\right)^2\sim \left({k_{\rm UV}}/{k_{\rm reh}}\right)^4$ \cite{Inomata:2019ivs}, where ${\cal H}$ is the conformal Hubble parameter. The additional factor $({k_{\rm reh}}/{k_{\rm UV}})$ originates from the fact that integral's contribution is dominated by a very narrow region, the width of which is set by this ratio. Here, we have also employed that fluctuations on scales much smaller than the typical transition time scale are affected by the finite width of the transition \cite{Inomata:2019ivs}. Due to radiation pressure, such small scale fluctuations are not efficiently produced which results in a suppression of $\Phi$ with respect to the exact instantaneous transition parametrised by $S_\Phi({k}/{k_{\rm reh}})$ \cite{Inomata:2019ivs}. We provide an explicit expression for $S_\Phi$ later. We also assumed that the highest contribution to the integral comes from the high momentum limit. This implies that the power spectrum must not fall too fast at small scales, which leads to consider only spectral indices $n<2$. With these assumptions and including the numerical coefficients, the amplitude of the peak GW emission frequency is 
\begin{align}\label{eq:gwspeak}
\Omega_{\rm GW,reh}^{\rm peak}\approx \frac{\pi A_\Phi^2}{9216\sqrt{3}} \left(\frac{4}{3}\right)^n S_\Phi^4({k}/{k_{\rm reh}})\left(\frac{k_{\rm UV}}{k_{\rm reh}}\right)^{7}
\end{align}
and so the contribution to stochastic GW background is
\begin{align}\label{eq:gwspeak}
\Omega_{\rm GW,reh}\approx \Omega_{\rm GW,reh}^{\rm peak}\,\left(\frac{k}{k_{\rm UV}}\right)^{7-2n}\Theta(k_{\rm UV}-k)~\,.
\end{align}
The subscript ``reh'' refers to the value of $\Omega_{\rm GW}$ right after reheating in the later radiation-dominated stage, where the energy density ratio of GWs and radiation stays constant. 
Note that by momentum conservation the GW spectrum has the cut-off at $k\sim 2k_{\rm UV}$. However, the peak is at $k\sim k_{\rm UV}$ and then decays rapidly \cite{Inomata:2020lmk,Domenech:2020ssp} (see Fig.~\ref{fig:neffGW}). Thus, for simplicity, we cut the GW spectrum at $k\sim k_{\rm UV}$.

We now consider induced GW production in the case of evaporating light PBHs. For simplicity, we assume that PBHs form at a particular time, with a monochromatic mass spectrum, and are randomly distributed across the universe with a given initial number density $n_{\rm PBH,i}$. This implies that the initial distribution of PBHs follows Poisson statistics and their (dimensionless) spectrum of fluctuations on small coarse-grained scales is \cite{Papanikolaou:2020qtd}
\begin{align}
{\cal P}_{{\rm PBH},i}(k)=\frac{2}{3\pi} \left(\frac{k}{k_{\rm UV}}\right)^3\,,
\end{align}
where the small scale cut-off is given in terms of the mean (comoving) distance between PBHs, that is
\begin{align}\label{eq:kuv}
k_{\rm UV}=\left(\frac{3}{4\pi}\frac{a_{i}^3}{n_{{\rm PBH},i}}\right)^{1/3}~,
\end{align}
where $a_i$ is the scale factor at the initial time.
Then the gravitational potential power spectrum of fluctuations can be estimated as
\begin{align}\label{eq:Pphi2}
{\cal P}_\Phi(k)={\cal P}_{{\rm PBH},i}(k){\cal T}_\Phi^2(k)\,,
\end{align}
where ${\cal T}_\Phi(k)$ is a transfer function that connects the PBH formation stage, set in an arbitrary cosmological scenario, to the PBH domination regime.  Let us consider the simplest scenario of PBH formation during an earlier radiation-dominated epoch. Then, the transfer function reads as \cite{Kodama:1986fg,Kodama:1986ud}
\begin{align}
{\cal T}_\Phi(k)=\left\{
\begin{aligned}
&\frac{1}{5}\qquad &k\ll k_{\rm eq}\\
&\frac{3}{4}\left(\frac{k_{\rm eq}}{k}\right)^2\qquad &k\gg k_{\rm eq}
\end{aligned}
\right.\,,
\end{align}
where $k_{\rm eq}=a_{\rm eq}H_{\rm eq}$ is the comoving horizon scale corresponding to radiation-matter (PBH) equality. Note that scales smaller that $k_{\rm eq}$ are suppressed due to the fact that a mode $k$ of $\Phi$ decays once it is inside the horizon. Some time after equality, the power spectrum of Eq.~\eqref{eq:Pphi2} is given by
\begin{align}\label{eq:Aphi}
{\cal P}_\Phi(k\gg k_{\rm eq})=\frac{3}{8\pi}\left(\frac{k_{\rm eq}}{k_{\rm UV}}\right)^4\left(\frac{k_{\rm UV}}{k}\right)\Theta\left(k_{\rm UV}-k\right)\,.
\end{align}
In the scenario where PBHs form in a matter-dominated era previous to the radiation-dominated era, there will be
 an additional scale $k_{\rm eq,\rm eRD}$ that corresponds to the comoving horizon scale when the universe enters the early radiation dominated stage,
 and an intermediate step needs to be considered in the transfer function above. 
However, the transfer function remains the same on sufficiently small scales $k\gg k_{\rm eq,\rm eRD}$, which are the scales of our interest. 

In a PBH-dominated stage, the reheating time and duration of the transition are set by BH evaporation. It then follows that (for details see Appendix~\ref{app:gws})
\begin{align}\label{eq:relatiosk2}
\frac{k_{\rm UV}}{k_{\rm reh}}\approx {2.3}\times 10^6
 \left(\frac{t_{\rm eva}(a_*)}{t_{\rm eva}(0)}\right)^{1/3}\left(\frac{M_{{\rm PBH},f}}{10^4{\rm g}}\right)^{2/3}\,.
\end{align}
Furthermore, following the results of \cite{Inomata:2020lmk} we have that
\begin{align}
S(k/k_{\rm reh})\approx\left(\sqrt{\frac{2}{3}}\frac{k}{k_{\rm reh}}\right)^{-1/3} \,.
\end{align}
Inserting equations \eqref{eq:Aphi} and \eqref{eq:relatiosk2} to \eqref{eq:gwspeak}, the peak of the GW spectrum (right after evaporation), at $k \sim k_{\rm UV}$, is approximately
\begin{align}\label{eq:gwspeak2}
&\Omega_{\rm GW,reh}(k_{\rm UV})
\\ &\approx  \frac{1}{49152\pi\sqrt{3}}\left({\frac{3}{2}}\right)^{2/3}\left(\frac{k_{\rm UV}}{k_{\rm reh}}\right)^{17/3} \left(\frac{k_{\rm eq}}{k_{\rm UV}}\right)^8
 \\ &\approx~8.3\times 10^{-3}\left(\frac{{\cal P}_\Phi}{10^{-23}}\right)^{2}
 \left(\frac{t_{\rm eva}(a_*)}{t_{\rm eva}(0)}\right)^{17/9}
\left(\frac{M_{{\rm PBH},f}}{10^7{\rm g}}\right)^{34/9}\,. \notag
\end{align}
The density fluctuations on $k\sim k_{\rm UV}$ become $O(1)$ at the time of reheating and effects of non-linearities could possibly become relevant. Hence, our results are approximate. This can be further improved with future numerical studies. On can then relate this result to the energy density ratio of gravitational waves at the time of BBN by \cite{Caprini:2018mtu}
\begin{align}
\Omega_{\rm GW,BBN}\approx&~ 0.39\left(\frac{g_*(T_{\rm reh})}{106.75}\right)^{-1/3}\Omega_{\rm GW,reh}( k_{\rm UV})\notag\\
=&~\frac{7}{8}\left(\frac{4}{11}\right)^{4/3}\Delta N_{\rm eff}\,,
\end{align}
where $\Delta N_{\rm eff}$ is
the effective number of neutrino species
and where we took into account the temperature dependence of the effective degrees of freedom $g_*(T)$ in the total energy density of radiation. Hence, we can write 
\begin{align}
\Delta N_{\rm eff}\approx&~ 1.4\times10^{-2}\left(\frac{{\cal P}_\Phi}{10^{-23}}\right)^{2}\left(\frac{t_{\rm eva}(a_*)}{t_{\rm eva}(0)}\right)^{17/9} \notag\\
&\times\left(\frac{g_*(T_{\rm reh})}{106.75}\right)^{-1/3}\left(\frac{M_{{\rm PBH},f}}{10^7{\rm g}}\right)^{34/9}\,.
\end{align}
We note that $g_*$ is not significantly affected by PBH spin~(e.g.~\cite{Masina:2021zpu}). The effect of PBH spin\footnote{We do not consider the special case of extremal BHs, which could modify phenomenology (e.g.~\cite{Bai:2019zcd}).} will be predominantly through the evaporation lifetime $t_{\rm eva}$, where for Kerr BH with Kerr spin parameter $a_{\ast} = 0.9999$ one has $t_{\rm eva}(a_*)/t_{\rm eva}(0) \sim 0.5$~\cite{Dong:2015yjs,Arbey:2019jmj}. To estimate the GW spectral density measured as a stochastic GW background by GW detectors, we just relate the GW density fraction to the total radiation density fraction by
\begin{align}
\Omega_{\rm GW,0}h^2\approx&~ 0.39\,\Omega_{r,0}h^2\left(\frac{g_*(T_{\rm reh})}{106.75}\right)^{-1/3}\Omega_{\rm GW,reh}\,,
\end{align}
where $\Omega_{r,0}h^2\approx 4.18 \times 10^{-5}$ is the present radiation energy density fraction~\cite{Aghanim:2018eyx}.

From Fig.~\ref{fig:neffGW} we see that PBH spin has a more dramatic effect on $\Delta N_{\rm eff}$ due to DR from PBH decay products compared to induced GWs. On the other hand, the induced GWs can lead to observable signatures even when PBHs are non-rotating. Further, $\Delta N_{\rm eff}$ from induced GWs is significantly affected by changes in PBH mass, in contrast to PBH decay emission.

While in the above we have considered a monochromatic PBH spectrum, formation models can often result in PBHs with distribution in spins and masses. While distribution in spins will significantly affect PBH decay signatures, this is not the case for induced GWs. This can be understood as follows, even for the case of Kerr BHs the lifetime will only change by a factor of $\sim 0.5$. On the other hand, distribution in masses can significantly modify the PBH lifetime that goes as $t_{\rm eva} \propto M_{\rm PBH}^3$ and hence the resulting induced GW signatures. The reason is that distribution in PBH lifetimes will result in a non-instantaneous effective evaporation as we considered above, which translates into more gradual deposits of evaporation radiation and less dramatic oscillations of the perturbations and suppression of GW production \cite{Inomata:2019zqy,Inomata:2020lmk}. A direct detection of stochastic GW counterpart signal (see Fig.~\ref{fig:neffGW}) would be characteristic of a narrow-width PBH mass-function.

As shown on the right of Fig.~\ref{fig:neffGW}, the $\Delta N_{\rm eff}$ signature can be testable and correlated further with the stochastic GW signals observed in GW experiments. This allows for further discrimination of the sources of the signals.
 
In conclusion, we have shown how coincidence signals associated with Hawking evaporation particle production as well as induced GWs can be used to explore and discriminate between different formation mechanisms for spinning and non-spinning evaporationg PBHs. This establishes a concrete methodology for charting the unexplored PBHs with masses $\lesssim 10^9$~g.
Our results are general and can be readily applied to a wide class of models.

\acknowledgments

We thank Kazunori Kohri for comments. G.D. as a Fellini fellow is supported by the European Union’s Horizon 2020 research and innovation programme under the Marie Sk{\l}odowska-Curie grant agreement No 754496. This work is supported in part by the JSPS KAKENHI Nos.~19H01895 and 20H04727. M.S. and V.T. are also supported by the World Premier International Research Center Initiative (WPI), MEXT, Japan.

\appendix

\section{Gravitational Wave Production}
\label{app:gws}

Here we present the details of the formulation if PBHs form by the collapse of large primordial fluctuations in a radiation dominated universe. Let us assume that primordial fluctuations have a sharp peak at some arbitrary scale $k_{ i}$. This scale enters the horizon at a time when $a_{ i}H_{ i}=k_i$ and PBHs form with a monochromatic mass-spectrum at 
\begin{align}
M_{{\rm PBH},i}=\frac{4\pi\gamma M_{\rm pl}^2}{H_i}\,.
\end{align}
In the above we assume that the mass enclosed inside the horizon collapses into a black hole with an efficiency parameter $\gamma$. In radiation domination, one has $\gamma\sim 0.2$~(see e.g.~\cite{Sasaki:2018dmp}). Starting with this initial time, PBHs start to evaporate by Hawking radiation.

\textit{Evaporation:} For a BH without spin, the evaporation rate goes as
\begin{align}\label{eq:hawkingevaporation}
\frac{dM_{\rm PBH}}{dt}=-\frac{A M_{\rm pl}^4}{M_{\rm PBH}^2}
\end{align}
where
\begin{align}
A=\frac{3.8\pi g_H(T_{\rm PBH})}{480}\approx 2.69 \left(\frac{g_H(T_{\rm PBH})}{108}\right) \,.
\end{align}
Hence, the evaporation time is
\begin{align}\label{eq:teva}
t_{\rm eva}\approx\frac{M_{{\rm PBH},f}^3}{3AM_{\rm pl}^4}\,.
\end{align}
If the PBH has spin, the evaporation is more efficient and the rate increases. We parametrise this with a spin dependence of the evaporation time, i.e. $t_{\rm eva}(a_*)$, where $a_*$ is the BH spin.
Assuming that PBH evaporate long after they dominate, the Hubble parameter at evaporation is given by 
\begin{align}
H_{\rm reh}=\frac{2}{3t_{\rm eva}}=\frac{\pi}{3\sqrt{10}M_{\rm pl}}g^{1/2}_*(T_{\rm reh})T^2_{\rm reh}\,,
\end{align}
where $g_*(T_{\rm reh})$ are the effective degrees of freedom of the energy density of the resulting radiation. The reheating (evaporation) temperature then reads
\begin{align}
T_{\rm reh}\approx&~2.76\times 10^4{\rm GeV} \left(\frac{M_{{\rm PBH},f}}{10^4{\rm g}}\right)^{-3/2}\nonumber\\ 
&\times
\left(\frac{t_{\rm eva}(a_*)}{t_{\rm eva}(0)}\right)^{-1/2}\left(\frac{g_*(T_{\rm reh})}{106.75}\right)^{-1/4}\,.
\end{align}
In order to have successful BBN, we roughly need thermal equilibrium at $T_{\rm reh}>4 \,{\rm MeV}$ that for a BH without (with maximal) spin imposes 
\begin{align}
M_{{\rm PBH},f}<5(7)\times10^8\,{\rm g}\,.
\end{align}
Note that this result depends on effective degrees of freedom at reheating $g_*(T_{\rm rh})$.

\textit{Density fluctuations:} PBH are produced very rarely. Only the largest primordial fluctuations from the tail of the distribution are above the necessary threshold. Thus, a good assumption is that PBHs are uniformly distributed across the universe with an initial mean PBH separation (assuming monochromatic spectrum) given by
\begin{align}
d_i\equiv\left(\frac{3M_{{\rm PBH},i}}{4\pi\rho_{{\rm PBH},i}}\right)^{1/3}=\gamma^{1/3}\beta^{-1/3}{H_i}^{-1}\,,
\end{align}
where in the last step we introduced $\beta$ as the initial density fraction of PBHs, i.e. $\beta\equiv\rho_{{\rm PBH},i}/\rho_{{\rm rad},i}$.
This uniform distribution leads to a Poisson spectrum for the density fluctuations on coarse grained scales given by \cite{Papanikolaou:2020qtd}
\begin{align}
\left\langle\frac{\delta\rho_{\rm PBH}(k)}{\rho_{\rm PBH}}\frac{\delta\rho_{\rm PBH}(k')}{{\rho_{\rm PBH}}}\right\rangle=\frac{4\pi}{3}\left(\frac{d}{a}\right)^3\,\delta(k+k')\,,
\end{align}
with a (comoving) cut-off given by
\begin{align}
k_{\rm UV}\equiv\frac{a_i}{d_i}=a_iH_i\beta^{1/3}\gamma^{-1/3}\,.
\end{align}
One can the find that the initial (isocurvature) dimensionless power spectrum is given by
\begin{align}
{\cal P}_{{\rm PBH},i}(k)=\frac{2}{3\pi} \left(\frac{k}{k_{\rm UV}}\right)^3\,.
\end{align}

\textit{From the initial stage to PBH domination:} At some moment, the initial fraction of PBH grows to dominate the universe. The time of radiation-PBH equality is approximately
\begin{align}
a_{\rm eq}/a_i=\beta^{-1}\,.
\end{align}
With such relation, we can compute all the relevant scales in terms of $\beta$ and the evaporation time by
\begin{align}\label{eq:relatiosk}
\frac{k_{\rm eq}}{k_i}={\sqrt{2}\beta}\,,\, \frac{k_{\rm UV}}{k_i}=\left(\frac{\beta}{\gamma}\right)^{1/3}\,,\, \frac{k_{\rm reh}}{k_i}=\left(\beta\frac{H_{\rm reh}}{H_i}\right)^{1/3}\,.
\end{align}
Using the transfer function presented in the main text, one finds that at some time inside the PBH domination, the initial isocurvature fluctuations are converted to curvature perturbations with an amplitude of the power spectrum at $k=k_{\rm UV}$ given by
\begin{align}
P_\Phi(k=k_{\rm UV})=\frac{3}{2\pi}\gamma^{4/3}\beta^{8/3}\,.
\end{align}
The resulting induced GW spectrum has a peak at $k\sim k_{\rm UV}$ with amplitude given by
\begin{align}\label{eq:gwspeak3}
&\Omega_{\rm GW,reh}(k_{\rm UV})
\\ &\approx  \frac{1}{49152\pi\sqrt{3}}\left({\frac{3}{2}}\right)^{2/3}\left(\frac{k_{\rm UV}}{k_{\rm reh}}\right)^{17/3} \left(\frac{k_{\rm eq}}{k_{\rm UV}}\right)^8
\nonumber\\ \approx&~5.5\times10^{-2}\left(\frac{\beta}{10^{-8}}\right)^{16/3}\left(\frac{t_{\rm eva}(a_*)}{t_{\rm eva}(0)}\right)^{17/9}
\left(\frac{M_{{\rm PBH},f}}{10^7{\rm g}}\right)^{34/9}\nonumber\,.
\end{align}
This is energy density of GWs at some time right after reheating. However, we must translate this results to the energy density of GWs today or at the time of BBN. To do that we must take into account the expansion of the universe and the change of the effective number of degrees of freedom from reheating to BBN. This is calculated by using that
\begin{align}
\frac{a_{\rm reh}}{a}=\frac{T}{T_{\rm reh}}\left(\frac{g_*(T)}{g_*(T_{\rm reh)}}\right)^{1/3}\,.
\end{align}
We then find that the density fraction of GWs at BBN is given by
\begin{align}
\Omega_{\rm GW,BBN}\approx&~ 0.39\,\left(\frac{g_*(T_{\rm reh})}{106.75}\right)^{-1/3}\Omega_{\rm GW,reh}\,.
\end{align}
From here one may extract the effective number of degrees of freedom, which yields
\begin{align}
\Delta N_{\rm eff}\approx&~ 0.1\left(\frac{\beta}{10^{-10}}\right)^{16/3}\left(\frac{t_{\rm eva}(a_*)}{t_{\rm eva}(0)}\right)^{17/9} \notag\\
&\times\left(\frac{g_*(T_{\rm reh})}{106.75}\right)^{-1/3}\left(\frac{M_{{\rm PBH},f}}{10^7{\rm g}}\right)^{34/9}\,.
\end{align}
To relate the amplitude at BBN to the amplitude measured by GW detectors, we can express it in terms of today's density parameter of radiation, $\Omega_{r,0}h^2$. In this case, the spectral density reads
\begin{align}
\Omega_{\rm GW,0}h^2\approx&~ 0.39\,\Omega_{r,0}h^2\left(\frac{g_*(T_{\rm reh})}{106.75}\right)^{-1/3}\Omega_{\rm GW,reh}\,.
\end{align}

\bibliography{bibliography}{}

 \newcommand{\noop}[1]{}
\begin{thebibliography}{91}%
\makeatletter
\providecommand \@ifxundefined [1]{%
 \@ifx{#1\undefined}
}%
\providecommand \@ifnum [1]{%
 \ifnum #1\expandafter \@firstoftwo
 \else \expandafter \@secondoftwo
 \fi
}%
\providecommand \@ifx [1]{%
 \ifx #1\expandafter \@firstoftwo
 \else \expandafter \@secondoftwo
 \fi
}%
\providecommand \natexlab [1]{#1}%
\providecommand \enquote  [1]{``#1''}%
\providecommand \bibnamefont  [1]{#1}%
\providecommand \bibfnamefont [1]{#1}%
\providecommand \citenamefont [1]{#1}%
\providecommand \href@noop [0]{\@secondoftwo}%
\providecommand \href [0]{\begingroup \@sanitize@url \@href}%
\providecommand \@href[1]{\@@startlink{#1}\@@href}%
\providecommand \@@href[1]{\endgroup#1\@@endlink}%
\providecommand \@sanitize@url [0]{\catcode `\\12\catcode `\$12\catcode
  `\&12\catcode `\#12\catcode `\^12\catcode `\_12\catcode `\%12\relax}%
\providecommand \@@startlink[1]{}%
\providecommand \@@endlink[0]{}%
\providecommand \url  [0]{\begingroup\@sanitize@url \@url }%
\providecommand \@url [1]{\endgroup\@href {#1}{\urlprefix }}%
\providecommand \urlprefix  [0]{URL }%
\providecommand \Eprint [0]{\href }%
\providecommand \doibase [0]{http://dx.doi.org/}%
\providecommand \selectlanguage [0]{\@gobble}%
\providecommand \bibinfo  [0]{\@secondoftwo}%
\providecommand \bibfield  [0]{\@secondoftwo}%
\providecommand \translation [1]{[#1]}%
\providecommand \BibitemOpen [0]{}%
\providecommand \bibitemStop [0]{}%
\providecommand \bibitemNoStop [0]{.\EOS\space}%
\providecommand \EOS [0]{\spacefactor3000\relax}%
\providecommand \BibitemShut  [1]{\csname bibitem#1\endcsname}%
\let\auto@bib@innerbib\@empty
%</preamble>
\bibitem [{\citenamefont {{Zel'dovich}}\ and\ \citenamefont
  {{Novikov}}(1967)}]{Zeldovich:1967}%
  \BibitemOpen
  \bibfield  {author} {\bibinfo {author} {\bibfnamefont {Y.~B.}\ \bibnamefont
  {{Zel'dovich}}}\ and\ \bibinfo {author} {\bibfnamefont {I.~D.}\ \bibnamefont
  {{Novikov}}},\ }\href@noop {} {\bibfield  {journal} {\bibinfo  {journal}
  {Sov. Astron.}\ }\textbf {\bibinfo {volume} {10}},\ \bibinfo {pages} {602}
  (\bibinfo {year} {1967})}\BibitemShut {NoStop}%
\bibitem [{\citenamefont {Hawking}(1971)}]{Hawking:1971ei}%
  \BibitemOpen
  \bibfield  {author} {\bibinfo {author} {\bibfnamefont {S.}~\bibnamefont
  {Hawking}},\ }\href@noop {} {\bibfield  {journal} {\bibinfo  {journal} {Mon.
  Not. Roy. Astron. Soc.}\ }\textbf {\bibinfo {volume} {152}},\ \bibinfo
  {pages} {75} (\bibinfo {year} {1971})}\BibitemShut {NoStop}%
%%CITATION = MNRAA,152,75;%%
\bibitem [{\citenamefont {Carr}\ and\ \citenamefont
  {Hawking}(1974)}]{Carr:1974nx}%
  \BibitemOpen
  \bibfield  {author} {\bibinfo {author} {\bibfnamefont {B.~J.}\ \bibnamefont
  {Carr}}\ and\ \bibinfo {author} {\bibfnamefont {S.~W.}\ \bibnamefont
  {Hawking}},\ }\href@noop {} {\bibfield  {journal} {\bibinfo  {journal} {Mon.
  Not. Roy. Astron. Soc.}\ }\textbf {\bibinfo {volume} {168}},\ \bibinfo
  {pages} {399} (\bibinfo {year} {1974})}\BibitemShut {NoStop}%
%%CITATION = MNRAA,168,399;%%
\bibitem [{\citenamefont {Chapline}(1975)}]{Chapline:1975ojl}%
  \BibitemOpen
  \bibfield  {author} {\bibinfo {author} {\bibfnamefont {G.~F.}\ \bibnamefont
  {Chapline}},\ }\href {\doibase 10.1038/253251a0} {\bibfield  {journal}
  {\bibinfo  {journal} {Nature}\ }\textbf {\bibinfo {volume} {253}},\ \bibinfo
  {pages} {251} (\bibinfo {year} {1975})}\BibitemShut {NoStop}%
\bibitem [{\citenamefont {Meszaros}(1975)}]{Meszaros:1975ef}%
  \BibitemOpen
  \bibfield  {author} {\bibinfo {author} {\bibfnamefont {P.}~\bibnamefont
  {Meszaros}},\ }\href@noop {} {\bibfield  {journal} {\bibinfo  {journal}
  {Astron. Astrophys.}\ }\textbf {\bibinfo {volume} {38}},\ \bibinfo {pages}
  {5} (\bibinfo {year} {1975})}\BibitemShut {NoStop}%
\bibitem [{\citenamefont {Carr}(1975)}]{Carr:1975qj}%
  \BibitemOpen
  \bibfield  {author} {\bibinfo {author} {\bibfnamefont {B.~J.}\ \bibnamefont
  {Carr}},\ }\href {\doibase 10.1086/153853} {\bibfield  {journal} {\bibinfo
  {journal} {Astrophys. J.}\ }\textbf {\bibinfo {volume} {201}},\ \bibinfo
  {pages} {1} (\bibinfo {year} {1975})}\BibitemShut {NoStop}%
\bibitem [{\citenamefont {Garcia-Bellido}\ \emph {et~al.}(1996)\citenamefont
  {Garcia-Bellido}, \citenamefont {Linde},\ and\ \citenamefont
  {Wands}}]{GarciaBellido:1996qt}%
  \BibitemOpen
  \bibfield  {author} {\bibinfo {author} {\bibfnamefont {J.}~\bibnamefont
  {Garcia-Bellido}}, \bibinfo {author} {\bibfnamefont {A.~D.}\ \bibnamefont
  {Linde}}, \ and\ \bibinfo {author} {\bibfnamefont {D.}~\bibnamefont
  {Wands}},\ }\href {\doibase 10.1103/PhysRevD.54.6040} {\bibfield  {journal}
  {\bibinfo  {journal} {Phys. Rev.}\ }\textbf {\bibinfo {volume} {D54}},\
  \bibinfo {pages} {6040} (\bibinfo {year} {1996})},\ \Eprint
  {http://arxiv.org/abs/astro-ph/9605094} {arXiv:astro-ph/9605094 [astro-ph]}
  \BibitemShut {NoStop}%
%%CITATION = ASTRO-PH/9605094;%%
\bibitem [{\citenamefont {Kawasaki}\ \emph {et~al.}(1998)\citenamefont
  {Kawasaki}, \citenamefont {Sugiyama},\ and\ \citenamefont
  {Yanagida}}]{Kawasaki:1997ju}%
  \BibitemOpen
  \bibfield  {author} {\bibinfo {author} {\bibfnamefont {M.}~\bibnamefont
  {Kawasaki}}, \bibinfo {author} {\bibfnamefont {N.}~\bibnamefont {Sugiyama}},
  \ and\ \bibinfo {author} {\bibfnamefont {T.}~\bibnamefont {Yanagida}},\
  }\href {\doibase 10.1103/PhysRevD.57.6050} {\bibfield  {journal} {\bibinfo
  {journal} {Phys. Rev. D}\ }\textbf {\bibinfo {volume} {57}},\ \bibinfo
  {pages} {6050} (\bibinfo {year} {1998})},\ \Eprint
  {http://arxiv.org/abs/hep-ph/9710259} {arXiv:hep-ph/9710259} \BibitemShut
  {NoStop}%
\bibitem [{\citenamefont {Khlopov}(2010)}]{Khlopov:2008qy}%
  \BibitemOpen
  \bibfield  {author} {\bibinfo {author} {\bibfnamefont {M.~{\relax Yu}.}\
  \bibnamefont {Khlopov}},\ }\href {\doibase 10.1088/1674-4527/10/6/001}
  {\bibfield  {journal} {\bibinfo  {journal} {Res. Astron. Astrophys.}\
  }\textbf {\bibinfo {volume} {10}},\ \bibinfo {pages} {495} (\bibinfo {year}
  {2010})},\ \Eprint {http://arxiv.org/abs/0801.0116} {arXiv:0801.0116
  [astro-ph]} \BibitemShut {NoStop}%
%%CITATION = ARXIV:0801.0116;%%
\bibitem [{\citenamefont {Frampton}\ \emph {et~al.}(2010)\citenamefont
  {Frampton}, \citenamefont {Kawasaki}, \citenamefont {Takahashi},\ and\
  \citenamefont {Yanagida}}]{Frampton:2010sw}%
  \BibitemOpen
  \bibfield  {author} {\bibinfo {author} {\bibfnamefont {P.~H.}\ \bibnamefont
  {Frampton}}, \bibinfo {author} {\bibfnamefont {M.}~\bibnamefont {Kawasaki}},
  \bibinfo {author} {\bibfnamefont {F.}~\bibnamefont {Takahashi}}, \ and\
  \bibinfo {author} {\bibfnamefont {T.~T.}\ \bibnamefont {Yanagida}},\ }\href
  {\doibase 10.1088/1475-7516/2010/04/023} {\bibfield  {journal} {\bibinfo
  {journal} {JCAP}\ }\textbf {\bibinfo {volume} {1004}},\ \bibinfo {pages}
  {023} (\bibinfo {year} {2010})},\ \Eprint {http://arxiv.org/abs/1001.2308}
  {arXiv:1001.2308 [hep-ph]} \BibitemShut {NoStop}%
%%CITATION = ARXIV:1001.2308;%%
\bibitem [{\citenamefont {Bird}\ \emph {et~al.}(2016)\citenamefont {Bird},
  \citenamefont {Cholis}, \citenamefont {Mu\~noz}, \citenamefont
  {Ali-Ha\"\i{}moud}, \citenamefont {Kamionkowski}, \citenamefont {Kovetz},
  \citenamefont {Raccanelli},\ and\ \citenamefont {Riess}}]{Bird:2016dcv}%
  \BibitemOpen
  \bibfield  {author} {\bibinfo {author} {\bibfnamefont {S.}~\bibnamefont
  {Bird}}, \bibinfo {author} {\bibfnamefont {I.}~\bibnamefont {Cholis}},
  \bibinfo {author} {\bibfnamefont {J.~B.}\ \bibnamefont {Mu\~noz}}, \bibinfo
  {author} {\bibfnamefont {Y.}~\bibnamefont {Ali-Ha\"\i{}moud}}, \bibinfo
  {author} {\bibfnamefont {M.}~\bibnamefont {Kamionkowski}}, \bibinfo {author}
  {\bibfnamefont {E.~D.}\ \bibnamefont {Kovetz}}, \bibinfo {author}
  {\bibfnamefont {A.}~\bibnamefont {Raccanelli}}, \ and\ \bibinfo {author}
  {\bibfnamefont {A.~G.}\ \bibnamefont {Riess}},\ }\href {\doibase
  10.1103/PhysRevLett.116.201301} {\bibfield  {journal} {\bibinfo  {journal}
  {Phys. Rev. Lett.}\ }\textbf {\bibinfo {volume} {116}},\ \bibinfo {pages}
  {201301} (\bibinfo {year} {2016})},\ \Eprint
  {http://arxiv.org/abs/1603.00464} {arXiv:1603.00464 [astro-ph.CO]}
  \BibitemShut {NoStop}%
\bibitem [{\citenamefont {Kawasaki}\ \emph {et~al.}(2016)\citenamefont
  {Kawasaki}, \citenamefont {Kusenko}, \citenamefont {Tada},\ and\
  \citenamefont {Yanagida}}]{Kawasaki:2016pql}%
  \BibitemOpen
  \bibfield  {author} {\bibinfo {author} {\bibfnamefont {M.}~\bibnamefont
  {Kawasaki}}, \bibinfo {author} {\bibfnamefont {A.}~\bibnamefont {Kusenko}},
  \bibinfo {author} {\bibfnamefont {Y.}~\bibnamefont {Tada}}, \ and\ \bibinfo
  {author} {\bibfnamefont {T.~T.}\ \bibnamefont {Yanagida}},\ }\href {\doibase
  10.1103/PhysRevD.94.083523} {\bibfield  {journal} {\bibinfo  {journal} {Phys.
  Rev.}\ }\textbf {\bibinfo {volume} {D94}},\ \bibinfo {pages} {083523}
  (\bibinfo {year} {2016})},\ \Eprint {http://arxiv.org/abs/1606.07631}
  {arXiv:1606.07631 [astro-ph.CO]} \BibitemShut {NoStop}%
%%CITATION = ARXIV:1606.07631;%%
\bibitem [{\citenamefont {Carr}\ \emph {et~al.}(2016)\citenamefont {Carr},
  \citenamefont {Kuhnel},\ and\ \citenamefont {Sandstad}}]{Carr:2016drx}%
  \BibitemOpen
  \bibfield  {author} {\bibinfo {author} {\bibfnamefont {B.}~\bibnamefont
  {Carr}}, \bibinfo {author} {\bibfnamefont {F.}~\bibnamefont {Kuhnel}}, \ and\
  \bibinfo {author} {\bibfnamefont {M.}~\bibnamefont {Sandstad}},\ }\href
  {\doibase 10.1103/PhysRevD.94.083504} {\bibfield  {journal} {\bibinfo
  {journal} {Phys. Rev.}\ }\textbf {\bibinfo {volume} {D94}},\ \bibinfo {pages}
  {083504} (\bibinfo {year} {2016})},\ \Eprint
  {http://arxiv.org/abs/1607.06077} {arXiv:1607.06077 [astro-ph.CO]}
  \BibitemShut {NoStop}%
%%CITATION = ARXIV:1607.06077;%%
\bibitem [{\citenamefont {Inomata}\ \emph
  {et~al.}(2017{\natexlab{a}})\citenamefont {Inomata}, \citenamefont
  {Kawasaki}, \citenamefont {Mukaida}, \citenamefont {Tada},\ and\
  \citenamefont {Yanagida}}]{Inomata:2016rbd}%
  \BibitemOpen
  \bibfield  {author} {\bibinfo {author} {\bibfnamefont {K.}~\bibnamefont
  {Inomata}}, \bibinfo {author} {\bibfnamefont {M.}~\bibnamefont {Kawasaki}},
  \bibinfo {author} {\bibfnamefont {K.}~\bibnamefont {Mukaida}}, \bibinfo
  {author} {\bibfnamefont {Y.}~\bibnamefont {Tada}}, \ and\ \bibinfo {author}
  {\bibfnamefont {T.~T.}\ \bibnamefont {Yanagida}},\ }\href {\doibase
  10.1103/PhysRevD.95.123510} {\bibfield  {journal} {\bibinfo  {journal} {Phys.
  Rev. D}\ }\textbf {\bibinfo {volume} {95}},\ \bibinfo {pages} {123510}
  (\bibinfo {year} {2017}{\natexlab{a}})},\ \Eprint
  {http://arxiv.org/abs/1611.06130} {arXiv:1611.06130 [astro-ph.CO]}
  \BibitemShut {NoStop}%
\bibitem [{\citenamefont {Pi}\ \emph {et~al.}(2018)\citenamefont {Pi},
  \citenamefont {Zhang}, \citenamefont {Huang},\ and\ \citenamefont
  {Sasaki}}]{Pi:2017gih}%
  \BibitemOpen
  \bibfield  {author} {\bibinfo {author} {\bibfnamefont {S.}~\bibnamefont
  {Pi}}, \bibinfo {author} {\bibfnamefont {Y.-l.}\ \bibnamefont {Zhang}},
  \bibinfo {author} {\bibfnamefont {Q.-G.}\ \bibnamefont {Huang}}, \ and\
  \bibinfo {author} {\bibfnamefont {M.}~\bibnamefont {Sasaki}},\ }\href
  {\doibase 10.1088/1475-7516/2018/05/042} {\bibfield  {journal} {\bibinfo
  {journal} {JCAP}\ }\textbf {\bibinfo {volume} {1805}},\ \bibinfo {pages}
  {042} (\bibinfo {year} {2018})},\ \Eprint {http://arxiv.org/abs/1712.09896}
  {arXiv:1712.09896 [astro-ph.CO]} \BibitemShut {NoStop}%
%%CITATION = ARXIV:1712.09896;%%
\bibitem [{\citenamefont {Inomata}\ \emph
  {et~al.}(2017{\natexlab{b}})\citenamefont {Inomata}, \citenamefont
  {Kawasaki}, \citenamefont {Mukaida}, \citenamefont {Tada},\ and\
  \citenamefont {Yanagida}}]{Inomata:2017okj}%
  \BibitemOpen
  \bibfield  {author} {\bibinfo {author} {\bibfnamefont {K.}~\bibnamefont
  {Inomata}}, \bibinfo {author} {\bibfnamefont {M.}~\bibnamefont {Kawasaki}},
  \bibinfo {author} {\bibfnamefont {K.}~\bibnamefont {Mukaida}}, \bibinfo
  {author} {\bibfnamefont {Y.}~\bibnamefont {Tada}}, \ and\ \bibinfo {author}
  {\bibfnamefont {T.~T.}\ \bibnamefont {Yanagida}},\ }\href {\doibase
  10.1103/PhysRevD.96.043504} {\bibfield  {journal} {\bibinfo  {journal} {Phys.
  Rev. D}\ }\textbf {\bibinfo {volume} {96}},\ \bibinfo {pages} {043504}
  (\bibinfo {year} {2017}{\natexlab{b}})},\ \Eprint
  {http://arxiv.org/abs/1701.02544} {arXiv:1701.02544 [astro-ph.CO]}
  \BibitemShut {NoStop}%
\bibitem [{\citenamefont {Garcia-Bellido}\ \emph {et~al.}(2017)\citenamefont
  {Garcia-Bellido}, \citenamefont {Peloso},\ and\ \citenamefont
  {Unal}}]{Garcia-Bellido:2017aan}%
  \BibitemOpen
  \bibfield  {author} {\bibinfo {author} {\bibfnamefont {J.}~\bibnamefont
  {Garcia-Bellido}}, \bibinfo {author} {\bibfnamefont {M.}~\bibnamefont
  {Peloso}}, \ and\ \bibinfo {author} {\bibfnamefont {C.}~\bibnamefont
  {Unal}},\ }\href {\doibase 10.1088/1475-7516/2017/09/013} {\bibfield
  {journal} {\bibinfo  {journal} {JCAP}\ }\textbf {\bibinfo {volume} {1709}},\
  \bibinfo {pages} {013} (\bibinfo {year} {2017})},\ \Eprint
  {http://arxiv.org/abs/1707.02441} {arXiv:1707.02441 [astro-ph.CO]}
  \BibitemShut {NoStop}%
%%CITATION = ARXIV:1707.02441;%%
\bibitem [{\citenamefont {Inoue}\ and\ \citenamefont
  {Kusenko}(2017)}]{Inoue:2017csr}%
  \BibitemOpen
  \bibfield  {author} {\bibinfo {author} {\bibfnamefont {Y.}~\bibnamefont
  {Inoue}}\ and\ \bibinfo {author} {\bibfnamefont {A.}~\bibnamefont
  {Kusenko}},\ }\href {\doibase 10.1088/1475-7516/2017/10/034} {\bibfield
  {journal} {\bibinfo  {journal} {JCAP}\ }\textbf {\bibinfo {volume} {1710}},\
  \bibinfo {pages} {034} (\bibinfo {year} {2017})},\ \Eprint
  {http://arxiv.org/abs/1705.00791} {arXiv:1705.00791 [astro-ph.CO]}
  \BibitemShut {NoStop}%
%%CITATION = ARXIV:1705.00791;%%
\bibitem [{\citenamefont {Georg}\ and\ \citenamefont
  {Watson}(2017)}]{Georg:2017mqk}%
  \BibitemOpen
  \bibfield  {author} {\bibinfo {author} {\bibfnamefont {J.}~\bibnamefont
  {Georg}}\ and\ \bibinfo {author} {\bibfnamefont {S.}~\bibnamefont {Watson}},\
  }\href {\doibase 10.1007/JHEP09(2017)138} {\bibfield  {journal} {\bibinfo
  {journal} {JHEP}\ }\textbf {\bibinfo {volume} {09}},\ \bibinfo {pages} {138}
  (\bibinfo {year} {2017})},\ \Eprint {http://arxiv.org/abs/1703.04825}
  {arXiv:1703.04825 [astro-ph.CO]} \BibitemShut {NoStop}%
\bibitem [{\citenamefont {Inomata}\ \emph
  {et~al.}(2017{\natexlab{c}})\citenamefont {Inomata}, \citenamefont
  {Kawasaki}, \citenamefont {Mukaida},\ and\ \citenamefont
  {Yanagida}}]{Inomata:2017bwi}%
  \BibitemOpen
  \bibfield  {author} {\bibinfo {author} {\bibfnamefont {K.}~\bibnamefont
  {Inomata}}, \bibinfo {author} {\bibfnamefont {M.}~\bibnamefont {Kawasaki}},
  \bibinfo {author} {\bibfnamefont {K.}~\bibnamefont {Mukaida}}, \ and\
  \bibinfo {author} {\bibfnamefont {T.~T.}\ \bibnamefont {Yanagida}},\
  }\href@noop {} {\  (\bibinfo {year} {2017}{\natexlab{c}})},\ \Eprint
  {http://arxiv.org/abs/1711.06129} {arXiv:1711.06129 [astro-ph.CO]}
  \BibitemShut {NoStop}%
%%CITATION = ARXIV:1711.06129;%%
\bibitem [{\citenamefont {Kocsis}\ \emph {et~al.}(2018)\citenamefont {Kocsis},
  \citenamefont {Suyama}, \citenamefont {Tanaka},\ and\ \citenamefont
  {Yokoyama}}]{Kocsis:2017yty}%
  \BibitemOpen
  \bibfield  {author} {\bibinfo {author} {\bibfnamefont {B.}~\bibnamefont
  {Kocsis}}, \bibinfo {author} {\bibfnamefont {T.}~\bibnamefont {Suyama}},
  \bibinfo {author} {\bibfnamefont {T.}~\bibnamefont {Tanaka}}, \ and\ \bibinfo
  {author} {\bibfnamefont {S.}~\bibnamefont {Yokoyama}},\ }\href {\doibase
  10.3847/1538-4357/aaa7f4} {\bibfield  {journal} {\bibinfo  {journal}
  {Astrophys. J.}\ }\textbf {\bibinfo {volume} {854}},\ \bibinfo {pages} {41}
  (\bibinfo {year} {2018})},\ \Eprint {http://arxiv.org/abs/1709.09007}
  {arXiv:1709.09007 [astro-ph.CO]} \BibitemShut {NoStop}%
\bibitem [{\citenamefont {Ando}\ \emph {et~al.}(2018)\citenamefont {Ando},
  \citenamefont {Inomata}, \citenamefont {Kawasaki}, \citenamefont {Mukaida},\
  and\ \citenamefont {Yanagida}}]{Ando:2017veq}%
  \BibitemOpen
  \bibfield  {author} {\bibinfo {author} {\bibfnamefont {K.}~\bibnamefont
  {Ando}}, \bibinfo {author} {\bibfnamefont {K.}~\bibnamefont {Inomata}},
  \bibinfo {author} {\bibfnamefont {M.}~\bibnamefont {Kawasaki}}, \bibinfo
  {author} {\bibfnamefont {K.}~\bibnamefont {Mukaida}}, \ and\ \bibinfo
  {author} {\bibfnamefont {T.~T.}\ \bibnamefont {Yanagida}},\ }\href {\doibase
  10.1103/PhysRevD.97.123512} {\bibfield  {journal} {\bibinfo  {journal} {Phys.
  Rev. D}\ }\textbf {\bibinfo {volume} {97}},\ \bibinfo {pages} {123512}
  (\bibinfo {year} {2018})},\ \Eprint {http://arxiv.org/abs/1711.08956}
  {arXiv:1711.08956 [astro-ph.CO]} \BibitemShut {NoStop}%
\bibitem [{\citenamefont {Niikura}\ \emph {et~al.}(2019)\citenamefont {Niikura}
  \emph {et~al.}}]{Niikura:2017zjd}%
  \BibitemOpen
  \bibfield  {author} {\bibinfo {author} {\bibfnamefont {H.}~\bibnamefont
  {Niikura}} \emph {et~al.},\ }\href {\doibase 10.1038/s41550-019-0723-1}
  {\bibfield  {journal} {\bibinfo  {journal} {Nat. Astron.}\ }\textbf {\bibinfo
  {volume} {3}},\ \bibinfo {pages} {524} (\bibinfo {year} {2019})},\ \Eprint
  {http://arxiv.org/abs/1701.02151} {arXiv:1701.02151 [astro-ph.CO]}
  \BibitemShut {NoStop}%
%%CITATION = ARXIV:1701.02151;%%
\bibitem [{\citenamefont {Cotner}\ and\ \citenamefont
  {Kusenko}(2017)}]{Cotner:2016cvr}%
  \BibitemOpen
  \bibfield  {author} {\bibinfo {author} {\bibfnamefont {E.}~\bibnamefont
  {Cotner}}\ and\ \bibinfo {author} {\bibfnamefont {A.}~\bibnamefont
  {Kusenko}},\ }\href {\doibase 10.1103/PhysRevLett.119.031103} {\bibfield
  {journal} {\bibinfo  {journal} {Phys. Rev. Lett.}\ }\textbf {\bibinfo
  {volume} {119}},\ \bibinfo {pages} {031103} (\bibinfo {year} {2017})},\
  \Eprint {http://arxiv.org/abs/1612.02529} {arXiv:1612.02529 [astro-ph.CO]}
  \BibitemShut {NoStop}%
%%CITATION = ARXIV:1612.02529;%%
\bibitem [{\citenamefont {Cotner}\ \emph {et~al.}(2019)\citenamefont {Cotner},
  \citenamefont {Kusenko}, \citenamefont {Sasaki},\ and\ \citenamefont
  {Takhistov}}]{Cotner:2019ykd}%
  \BibitemOpen
  \bibfield  {author} {\bibinfo {author} {\bibfnamefont {E.}~\bibnamefont
  {Cotner}}, \bibinfo {author} {\bibfnamefont {A.}~\bibnamefont {Kusenko}},
  \bibinfo {author} {\bibfnamefont {M.}~\bibnamefont {Sasaki}}, \ and\ \bibinfo
  {author} {\bibfnamefont {V.}~\bibnamefont {Takhistov}},\ }\href {\doibase
  10.1088/1475-7516/2019/10/077} {\bibfield  {journal} {\bibinfo  {journal}
  {JCAP}\ }\textbf {\bibinfo {volume} {1910}},\ \bibinfo {pages} {077}
  (\bibinfo {year} {2019})},\ \Eprint {http://arxiv.org/abs/1907.10613}
  {arXiv:1907.10613 [astro-ph.CO]} \BibitemShut {NoStop}%
%%CITATION = ARXIV:1907.10613;%%
\bibitem [{\citenamefont {Cotner}\ \emph {et~al.}(2018)\citenamefont {Cotner},
  \citenamefont {Kusenko},\ and\ \citenamefont {Takhistov}}]{Cotner:2018vug}%
  \BibitemOpen
  \bibfield  {author} {\bibinfo {author} {\bibfnamefont {E.}~\bibnamefont
  {Cotner}}, \bibinfo {author} {\bibfnamefont {A.}~\bibnamefont {Kusenko}}, \
  and\ \bibinfo {author} {\bibfnamefont {V.}~\bibnamefont {Takhistov}},\ }\href
  {\doibase 10.1103/PhysRevD.98.083513} {\bibfield  {journal} {\bibinfo
  {journal} {Phys. Rev.}\ }\textbf {\bibinfo {volume} {D98}},\ \bibinfo {pages}
  {083513} (\bibinfo {year} {2018})},\ \Eprint
  {http://arxiv.org/abs/1801.03321} {arXiv:1801.03321 [astro-ph.CO]}
  \BibitemShut {NoStop}%
%%CITATION = ARXIV:1801.03321;%%
\bibitem [{\citenamefont {Sasaki}\ \emph {et~al.}(2018)\citenamefont {Sasaki},
  \citenamefont {Suyama}, \citenamefont {Tanaka},\ and\ \citenamefont
  {Yokoyama}}]{Sasaki:2018dmp}%
  \BibitemOpen
  \bibfield  {author} {\bibinfo {author} {\bibfnamefont {M.}~\bibnamefont
  {Sasaki}}, \bibinfo {author} {\bibfnamefont {T.}~\bibnamefont {Suyama}},
  \bibinfo {author} {\bibfnamefont {T.}~\bibnamefont {Tanaka}}, \ and\ \bibinfo
  {author} {\bibfnamefont {S.}~\bibnamefont {Yokoyama}},\ }\href {\doibase
  10.1088/1361-6382/aaa7b4} {\bibfield  {journal} {\bibinfo  {journal} {Class.
  Quant. Grav.}\ }\textbf {\bibinfo {volume} {35}},\ \bibinfo {pages} {063001}
  (\bibinfo {year} {2018})},\ \Eprint {http://arxiv.org/abs/1801.05235}
  {arXiv:1801.05235 [astro-ph.CO]} \BibitemShut {NoStop}%
%%CITATION = ARXIV:1801.05235;%%
\bibitem [{\citenamefont {Carr}\ and\ \citenamefont
  {Silk}(2018)}]{Carr:2018rid}%
  \BibitemOpen
  \bibfield  {author} {\bibinfo {author} {\bibfnamefont {B.}~\bibnamefont
  {Carr}}\ and\ \bibinfo {author} {\bibfnamefont {J.}~\bibnamefont {Silk}},\
  }\href {\doibase 10.1093/mnras/sty1204} {\bibfield  {journal} {\bibinfo
  {journal} {Mon. Not. Roy. Astron. Soc.}\ }\textbf {\bibinfo {volume} {478}},\
  \bibinfo {pages} {3756} (\bibinfo {year} {2018})},\ \Eprint
  {http://arxiv.org/abs/1801.00672} {arXiv:1801.00672 [astro-ph.CO]}
  \BibitemShut {NoStop}%
%%CITATION = ARXIV:1801.00672;%%
\bibitem [{\citenamefont {Flores}\ and\ \citenamefont
  {Kusenko}(2020)}]{Flores:2020drq}%
  \BibitemOpen
  \bibfield  {author} {\bibinfo {author} {\bibfnamefont {M.~M.}\ \bibnamefont
  {Flores}}\ and\ \bibinfo {author} {\bibfnamefont {A.}~\bibnamefont
  {Kusenko}},\ }\href@noop {} {\  (\bibinfo {year} {2020})},\ \Eprint
  {http://arxiv.org/abs/2008.12456} {arXiv:2008.12456 [astro-ph.CO]}
  \BibitemShut {NoStop}%
\bibitem [{\citenamefont {Deng}\ and\ \citenamefont
  {Vilenkin}(2017)}]{Deng:2017uwc}%
  \BibitemOpen
  \bibfield  {author} {\bibinfo {author} {\bibfnamefont {H.}~\bibnamefont
  {Deng}}\ and\ \bibinfo {author} {\bibfnamefont {A.}~\bibnamefont
  {Vilenkin}},\ }\href {\doibase 10.1088/1475-7516/2017/12/044} {\bibfield
  {journal} {\bibinfo  {journal} {JCAP}\ }\textbf {\bibinfo {volume} {1712}},\
  \bibinfo {pages} {044} (\bibinfo {year} {2017})},\ \Eprint
  {http://arxiv.org/abs/1710.02865} {arXiv:1710.02865 [gr-qc]} \BibitemShut
  {NoStop}%
%%CITATION = ARXIV:1710.02865;%%
\bibitem [{\citenamefont {Kusenko}\ \emph
  {et~al.}(2020{\natexlab{a}})\citenamefont {Kusenko}, \citenamefont {Sasaki},
  \citenamefont {Sugiyama}, \citenamefont {Takada}, \citenamefont {Takhistov},\
  and\ \citenamefont {Vitagliano}}]{Kusenko:2020pcg}%
  \BibitemOpen
  \bibfield  {author} {\bibinfo {author} {\bibfnamefont {A.}~\bibnamefont
  {Kusenko}}, \bibinfo {author} {\bibfnamefont {M.}~\bibnamefont {Sasaki}},
  \bibinfo {author} {\bibfnamefont {S.}~\bibnamefont {Sugiyama}}, \bibinfo
  {author} {\bibfnamefont {M.}~\bibnamefont {Takada}}, \bibinfo {author}
  {\bibfnamefont {V.}~\bibnamefont {Takhistov}}, \ and\ \bibinfo {author}
  {\bibfnamefont {E.}~\bibnamefont {Vitagliano}},\ }\href@noop {} {\  (\bibinfo
  {year} {2020}{\natexlab{a}})},\ \Eprint {http://arxiv.org/abs/2001.09160}
  {arXiv:2001.09160 [astro-ph.CO]} \BibitemShut {NoStop}%
\bibitem [{\citenamefont {Lu}\ \emph {et~al.}(2020)\citenamefont {Lu},
  \citenamefont {Takhistov}, \citenamefont {Gelmini}, \citenamefont {Hayashi},
  \citenamefont {Inoue},\ and\ \citenamefont {Kusenko}}]{Lu:2020bmd}%
  \BibitemOpen
  \bibfield  {author} {\bibinfo {author} {\bibfnamefont {P.}~\bibnamefont
  {Lu}}, \bibinfo {author} {\bibfnamefont {V.}~\bibnamefont {Takhistov}},
  \bibinfo {author} {\bibfnamefont {G.~B.}\ \bibnamefont {Gelmini}}, \bibinfo
  {author} {\bibfnamefont {K.}~\bibnamefont {Hayashi}}, \bibinfo {author}
  {\bibfnamefont {Y.}~\bibnamefont {Inoue}}, \ and\ \bibinfo {author}
  {\bibfnamefont {A.}~\bibnamefont {Kusenko}},\ }\href@noop {} {\  (\bibinfo
  {year} {2020})},\ \Eprint {http://arxiv.org/abs/2007.02213} {arXiv:2007.02213
  [astro-ph.CO]} \BibitemShut {NoStop}%
\bibitem [{\citenamefont {Fuller}\ \emph {et~al.}(2017)\citenamefont {Fuller},
  \citenamefont {Kusenko},\ and\ \citenamefont {Takhistov}}]{Fuller:2017uyd}%
  \BibitemOpen
  \bibfield  {author} {\bibinfo {author} {\bibfnamefont {G.~M.}\ \bibnamefont
  {Fuller}}, \bibinfo {author} {\bibfnamefont {A.}~\bibnamefont {Kusenko}}, \
  and\ \bibinfo {author} {\bibfnamefont {V.}~\bibnamefont {Takhistov}},\ }\href
  {\doibase 10.1103/PhysRevLett.119.061101} {\bibfield  {journal} {\bibinfo
  {journal} {Phys. Rev. Lett.}\ }\textbf {\bibinfo {volume} {119}},\ \bibinfo
  {pages} {061101} (\bibinfo {year} {2017})},\ \Eprint
  {http://arxiv.org/abs/1704.01129} {arXiv:1704.01129 [astro-ph.HE]}
  \BibitemShut {NoStop}%
%%CITATION = ARXIV:1704.01129;%%
\bibitem [{\citenamefont {Kusenko}\ \emph
  {et~al.}(2020{\natexlab{b}})\citenamefont {Kusenko}, \citenamefont
  {Takhistov}, \citenamefont {Yamada},\ and\ \citenamefont
  {Yamazaki}}]{Kusenko:2019kcu}%
  \BibitemOpen
  \bibfield  {author} {\bibinfo {author} {\bibfnamefont {A.}~\bibnamefont
  {Kusenko}}, \bibinfo {author} {\bibfnamefont {V.}~\bibnamefont {Takhistov}},
  \bibinfo {author} {\bibfnamefont {M.}~\bibnamefont {Yamada}}, \ and\ \bibinfo
  {author} {\bibfnamefont {M.}~\bibnamefont {Yamazaki}},\ }\href {\doibase
  10.1016/j.physletb.2020.135369} {\bibfield  {journal} {\bibinfo  {journal}
  {Phys. Lett. B}\ }\textbf {\bibinfo {volume} {804}},\ \bibinfo {pages}
  {135369} (\bibinfo {year} {2020}{\natexlab{b}})},\ \Eprint
  {http://arxiv.org/abs/1908.10930} {arXiv:1908.10930 [hep-th]} \BibitemShut
  {NoStop}%
\bibitem [{\citenamefont {Takhistov}\ \emph
  {et~al.}(2021{\natexlab{a}})\citenamefont {Takhistov}, \citenamefont
  {Fuller},\ and\ \citenamefont {Kusenko}}]{Takhistov:2020vxs}%
  \BibitemOpen
  \bibfield  {author} {\bibinfo {author} {\bibfnamefont {V.}~\bibnamefont
  {Takhistov}}, \bibinfo {author} {\bibfnamefont {G.~M.}\ \bibnamefont
  {Fuller}}, \ and\ \bibinfo {author} {\bibfnamefont {A.}~\bibnamefont
  {Kusenko}},\ }\href {\doibase 10.1103/PhysRevLett.126.071101} {\bibfield
  {journal} {\bibinfo  {journal} {Phys. Rev. Lett.}\ }\textbf {\bibinfo
  {volume} {126}},\ \bibinfo {pages} {071101} (\bibinfo {year}
  {2021}{\natexlab{a}})},\ \Eprint {http://arxiv.org/abs/2008.12780}
  {arXiv:2008.12780 [astro-ph.HE]} \BibitemShut {NoStop}%
\bibitem [{\citenamefont {Kawasaki}\ and\ \citenamefont
  {Takhistov}(2018)}]{Kawasaki:2018daf}%
  \BibitemOpen
  \bibfield  {author} {\bibinfo {author} {\bibfnamefont {M.}~\bibnamefont
  {Kawasaki}}\ and\ \bibinfo {author} {\bibfnamefont {V.}~\bibnamefont
  {Takhistov}},\ }\href {\doibase 10.1103/PhysRevD.98.123514} {\bibfield
  {journal} {\bibinfo  {journal} {Phys. Rev.}\ }\textbf {\bibinfo {volume}
  {D98}},\ \bibinfo {pages} {123514} (\bibinfo {year} {2018})},\ \Eprint
  {http://arxiv.org/abs/1810.02547} {arXiv:1810.02547 [hep-th]} \BibitemShut
  {NoStop}%
%%CITATION = ARXIV:1810.02547;%%
\bibitem [{\citenamefont {Takhistov}(2018)}]{Takhistov:2017bpt}%
  \BibitemOpen
  \bibfield  {author} {\bibinfo {author} {\bibfnamefont {V.}~\bibnamefont
  {Takhistov}},\ }\href {\doibase 10.1016/j.physletb.2018.05.026} {\bibfield
  {journal} {\bibinfo  {journal} {Phys. Lett.}\ }\textbf {\bibinfo {volume}
  {B782}},\ \bibinfo {pages} {77} (\bibinfo {year} {2018})},\ \Eprint
  {http://arxiv.org/abs/1707.05849} {arXiv:1707.05849 [astro-ph.CO]}
  \BibitemShut {NoStop}%
%%CITATION = ARXIV:1707.05849;%%
\bibitem [{\citenamefont {Takhistov}(2019)}]{Takhistov:2017nmt}%
  \BibitemOpen
  \bibfield  {author} {\bibinfo {author} {\bibfnamefont {V.}~\bibnamefont
  {Takhistov}},\ }\href {\doibase 10.1016/j.physletb.2018.12.043} {\bibfield
  {journal} {\bibinfo  {journal} {Phys. Lett.}\ }\textbf {\bibinfo {volume}
  {B789}},\ \bibinfo {pages} {538} (\bibinfo {year} {2019})},\ \Eprint
  {http://arxiv.org/abs/1710.09458} {arXiv:1710.09458 [astro-ph.HE]}
  \BibitemShut {NoStop}%
%%CITATION = ARXIV:1710.09458;%%
\bibitem [{\citenamefont {Laha}\ \emph {et~al.}(2020)\citenamefont {Laha},
  \citenamefont {Lu},\ and\ \citenamefont {Takhistov}}]{Laha:2020vhg}%
  \BibitemOpen
  \bibfield  {author} {\bibinfo {author} {\bibfnamefont {R.}~\bibnamefont
  {Laha}}, \bibinfo {author} {\bibfnamefont {P.}~\bibnamefont {Lu}}, \ and\
  \bibinfo {author} {\bibfnamefont {V.}~\bibnamefont {Takhistov}},\ }\href@noop
  {} {\  (\bibinfo {year} {2020})},\ \Eprint {http://arxiv.org/abs/2009.11837}
  {arXiv:2009.11837 [astro-ph.CO]} \BibitemShut {NoStop}%
\bibitem [{\citenamefont {Sugiyama}\ \emph {et~al.}(2021)\citenamefont
  {Sugiyama}, \citenamefont {Takhistov}, \citenamefont {Vitagliano},
  \citenamefont {Kusenko}, \citenamefont {Sasaki},\ and\ \citenamefont
  {Takada}}]{Sugiyama:2020roc}%
  \BibitemOpen
  \bibfield  {author} {\bibinfo {author} {\bibfnamefont {S.}~\bibnamefont
  {Sugiyama}}, \bibinfo {author} {\bibfnamefont {V.}~\bibnamefont {Takhistov}},
  \bibinfo {author} {\bibfnamefont {E.}~\bibnamefont {Vitagliano}}, \bibinfo
  {author} {\bibfnamefont {A.}~\bibnamefont {Kusenko}}, \bibinfo {author}
  {\bibfnamefont {M.}~\bibnamefont {Sasaki}}, \ and\ \bibinfo {author}
  {\bibfnamefont {M.}~\bibnamefont {Takada}},\ }\href {\doibase
  10.1016/j.physletb.2021.136097} {\bibfield  {journal} {\bibinfo  {journal}
  {Phys. Lett. B}\ }\textbf {\bibinfo {volume} {814}},\ \bibinfo {pages}
  {136097} (\bibinfo {year} {2021})},\ \Eprint
  {http://arxiv.org/abs/2010.02189} {arXiv:2010.02189 [astro-ph.CO]}
  \BibitemShut {NoStop}%
\bibitem [{\citenamefont {Takhistov}\ \emph
  {et~al.}(2021{\natexlab{b}})\citenamefont {Takhistov}, \citenamefont {Lu},
  \citenamefont {Gelmini}, \citenamefont {Hayashi}, \citenamefont {Inoue},\
  and\ \citenamefont {Kusenko}}]{Takhistov:2021aqx}%
  \BibitemOpen
  \bibfield  {author} {\bibinfo {author} {\bibfnamefont {V.}~\bibnamefont
  {Takhistov}}, \bibinfo {author} {\bibfnamefont {P.}~\bibnamefont {Lu}},
  \bibinfo {author} {\bibfnamefont {G.~B.}\ \bibnamefont {Gelmini}}, \bibinfo
  {author} {\bibfnamefont {K.}~\bibnamefont {Hayashi}}, \bibinfo {author}
  {\bibfnamefont {Y.}~\bibnamefont {Inoue}}, \ and\ \bibinfo {author}
  {\bibfnamefont {A.}~\bibnamefont {Kusenko}},\ }\href@noop {} {\  (\bibinfo
  {year} {2021}{\natexlab{b}})},\ \Eprint {http://arxiv.org/abs/2105.06099}
  {arXiv:2105.06099 [astro-ph.GA]} \BibitemShut {NoStop}%
\bibitem [{\citenamefont {Hawking}(1975)}]{Hawking:1974sw}%
  \BibitemOpen
  \bibfield  {author} {\bibinfo {author} {\bibfnamefont {S.}~\bibnamefont
  {Hawking}},\ }\href {\doibase 10.1007/BF02345020} {\bibfield  {journal}
  {\bibinfo  {journal} {Commun. Math. Phys.}\ }\textbf {\bibinfo {volume}
  {43}},\ \bibinfo {pages} {199} (\bibinfo {year} {1975})},\ \bibinfo {note}
  {[Erratum: Commun.Math.Phys. 46, 206 (1976)]}\BibitemShut {NoStop}%
\bibitem [{\citenamefont {Page}(1976{\natexlab{a}})}]{Page:1976df}%
  \BibitemOpen
  \bibfield  {author} {\bibinfo {author} {\bibfnamefont {D.~N.}\ \bibnamefont
  {Page}},\ }\href {\doibase 10.1103/PhysRevD.13.198} {\bibfield  {journal}
  {\bibinfo  {journal} {Phys. Rev. D}\ }\textbf {\bibinfo {volume} {13}},\
  \bibinfo {pages} {198} (\bibinfo {year} {1976}{\natexlab{a}})}\BibitemShut
  {NoStop}%
\bibitem [{\citenamefont {Page}(1977)}]{Page:1977um}%
  \BibitemOpen
  \bibfield  {author} {\bibinfo {author} {\bibfnamefont {D.~N.}\ \bibnamefont
  {Page}},\ }\href {\doibase 10.1103/PhysRevD.16.2402} {\bibfield  {journal}
  {\bibinfo  {journal} {Phys. Rev. D}\ }\textbf {\bibinfo {volume} {16}},\
  \bibinfo {pages} {2402} (\bibinfo {year} {1977})}\BibitemShut {NoStop}%
\bibitem [{\citenamefont {Fujita}\ \emph {et~al.}(2014)\citenamefont {Fujita},
  \citenamefont {Kawasaki}, \citenamefont {Harigaya},\ and\ \citenamefont
  {Matsuda}}]{Fujita:2014hha}%
  \BibitemOpen
  \bibfield  {author} {\bibinfo {author} {\bibfnamefont {T.}~\bibnamefont
  {Fujita}}, \bibinfo {author} {\bibfnamefont {M.}~\bibnamefont {Kawasaki}},
  \bibinfo {author} {\bibfnamefont {K.}~\bibnamefont {Harigaya}}, \ and\
  \bibinfo {author} {\bibfnamefont {R.}~\bibnamefont {Matsuda}},\ }\href
  {\doibase 10.1103/PhysRevD.89.103501} {\bibfield  {journal} {\bibinfo
  {journal} {Phys. Rev. D}\ }\textbf {\bibinfo {volume} {89}},\ \bibinfo
  {pages} {103501} (\bibinfo {year} {2014})},\ \Eprint
  {http://arxiv.org/abs/1401.1909} {arXiv:1401.1909 [astro-ph.CO]} \BibitemShut
  {NoStop}%
\bibitem [{\citenamefont {Lennon}\ \emph {et~al.}(2018)\citenamefont {Lennon},
  \citenamefont {March-Russell}, \citenamefont {Petrossian-Byrne},\ and\
  \citenamefont {Tillim}}]{Lennon:2017tqq}%
  \BibitemOpen
  \bibfield  {author} {\bibinfo {author} {\bibfnamefont {O.}~\bibnamefont
  {Lennon}}, \bibinfo {author} {\bibfnamefont {J.}~\bibnamefont
  {March-Russell}}, \bibinfo {author} {\bibfnamefont {R.}~\bibnamefont
  {Petrossian-Byrne}}, \ and\ \bibinfo {author} {\bibfnamefont
  {H.}~\bibnamefont {Tillim}},\ }\href {\doibase 10.1088/1475-7516/2018/04/009}
  {\bibfield  {journal} {\bibinfo  {journal} {JCAP}\ }\textbf {\bibinfo
  {volume} {04}},\ \bibinfo {pages} {009} (\bibinfo {year} {2018})},\ \Eprint
  {http://arxiv.org/abs/1712.07664} {arXiv:1712.07664 [hep-ph]} \BibitemShut
  {NoStop}%
\bibitem [{\citenamefont {Morrison}\ \emph {et~al.}(2019)\citenamefont
  {Morrison}, \citenamefont {Profumo},\ and\ \citenamefont
  {Yu}}]{Morrison:2018xla}%
  \BibitemOpen
  \bibfield  {author} {\bibinfo {author} {\bibfnamefont {L.}~\bibnamefont
  {Morrison}}, \bibinfo {author} {\bibfnamefont {S.}~\bibnamefont {Profumo}}, \
  and\ \bibinfo {author} {\bibfnamefont {Y.}~\bibnamefont {Yu}},\ }\href
  {\doibase 10.1088/1475-7516/2019/05/005} {\bibfield  {journal} {\bibinfo
  {journal} {JCAP}\ }\textbf {\bibinfo {volume} {05}},\ \bibinfo {pages} {005}
  (\bibinfo {year} {2019})},\ \Eprint {http://arxiv.org/abs/1812.10606}
  {arXiv:1812.10606 [astro-ph.CO]} \BibitemShut {NoStop}%
\bibitem [{\citenamefont {Masina}(2020)}]{Masina:2020xhk}%
  \BibitemOpen
  \bibfield  {author} {\bibinfo {author} {\bibfnamefont {I.}~\bibnamefont
  {Masina}},\ }\href {\doibase 10.1140/epjp/s13360-020-00564-9} {\bibfield
  {journal} {\bibinfo  {journal} {Eur. Phys. J. Plus}\ }\textbf {\bibinfo
  {volume} {135}},\ \bibinfo {pages} {552} (\bibinfo {year} {2020})},\ \Eprint
  {http://arxiv.org/abs/2004.04740} {arXiv:2004.04740 [hep-ph]} \BibitemShut
  {NoStop}%
\bibitem [{\citenamefont {Hooper}\ \emph {et~al.}(2019)\citenamefont {Hooper},
  \citenamefont {Krnjaic},\ and\ \citenamefont {McDermott}}]{Hooper:2019gtx}%
  \BibitemOpen
  \bibfield  {author} {\bibinfo {author} {\bibfnamefont {D.}~\bibnamefont
  {Hooper}}, \bibinfo {author} {\bibfnamefont {G.}~\bibnamefont {Krnjaic}}, \
  and\ \bibinfo {author} {\bibfnamefont {S.~D.}\ \bibnamefont {McDermott}},\
  }\href {\doibase 10.1007/JHEP08(2019)001} {\bibfield  {journal} {\bibinfo
  {journal} {JHEP}\ }\textbf {\bibinfo {volume} {08}},\ \bibinfo {pages} {001}
  (\bibinfo {year} {2019})},\ \Eprint {http://arxiv.org/abs/1905.01301}
  {arXiv:1905.01301 [hep-ph]} \BibitemShut {NoStop}%
\bibitem [{\citenamefont {Hooper}\ \emph {et~al.}(2020)\citenamefont {Hooper},
  \citenamefont {Krnjaic}, \citenamefont {March-Russell}, \citenamefont
  {McDermott},\ and\ \citenamefont {Petrossian-Byrne}}]{Hooper:2020evu}%
  \BibitemOpen
  \bibfield  {author} {\bibinfo {author} {\bibfnamefont {D.}~\bibnamefont
  {Hooper}}, \bibinfo {author} {\bibfnamefont {G.}~\bibnamefont {Krnjaic}},
  \bibinfo {author} {\bibfnamefont {J.}~\bibnamefont {March-Russell}}, \bibinfo
  {author} {\bibfnamefont {S.~D.}\ \bibnamefont {McDermott}}, \ and\ \bibinfo
  {author} {\bibfnamefont {R.}~\bibnamefont {Petrossian-Byrne}},\ }\href@noop
  {} {\  (\bibinfo {year} {2020})},\ \Eprint {http://arxiv.org/abs/2004.00618}
  {arXiv:2004.00618 [astro-ph.CO]} \BibitemShut {NoStop}%
\bibitem [{\citenamefont {Lunardini}\ and\ \citenamefont
  {Perez-Gonzalez}(2020)}]{Lunardini:2019zob}%
  \BibitemOpen
  \bibfield  {author} {\bibinfo {author} {\bibfnamefont {C.}~\bibnamefont
  {Lunardini}}\ and\ \bibinfo {author} {\bibfnamefont {Y.~F.}\ \bibnamefont
  {Perez-Gonzalez}},\ }\href {\doibase 10.1088/1475-7516/2020/08/014}
  {\bibfield  {journal} {\bibinfo  {journal} {JCAP}\ }\textbf {\bibinfo
  {volume} {08}},\ \bibinfo {pages} {014} (\bibinfo {year} {2020})},\ \Eprint
  {http://arxiv.org/abs/1910.07864} {arXiv:1910.07864 [hep-ph]} \BibitemShut
  {NoStop}%
\bibitem [{\citenamefont {Gondolo}\ \emph {et~al.}(2020)\citenamefont
  {Gondolo}, \citenamefont {Sandick},\ and\ \citenamefont {Shams
  Es~Haghi}}]{Gondolo:2020uqv}%
  \BibitemOpen
  \bibfield  {author} {\bibinfo {author} {\bibfnamefont {P.}~\bibnamefont
  {Gondolo}}, \bibinfo {author} {\bibfnamefont {P.}~\bibnamefont {Sandick}}, \
  and\ \bibinfo {author} {\bibfnamefont {B.}~\bibnamefont {Shams Es~Haghi}},\
  }\href {\doibase 10.1103/PhysRevD.102.095018} {\bibfield  {journal} {\bibinfo
   {journal} {Phys. Rev. D}\ }\textbf {\bibinfo {volume} {102}},\ \bibinfo
  {pages} {095018} (\bibinfo {year} {2020})},\ \Eprint
  {http://arxiv.org/abs/2009.02424} {arXiv:2009.02424 [hep-ph]} \BibitemShut
  {NoStop}%
\bibitem [{\citenamefont {Carr}\ \emph {et~al.}(2020)\citenamefont {Carr},
  \citenamefont {Kohri}, \citenamefont {Sendouda},\ and\ \citenamefont
  {Yokoyama}}]{Carr:2020gox}%
  \BibitemOpen
  \bibfield  {author} {\bibinfo {author} {\bibfnamefont {B.}~\bibnamefont
  {Carr}}, \bibinfo {author} {\bibfnamefont {K.}~\bibnamefont {Kohri}},
  \bibinfo {author} {\bibfnamefont {Y.}~\bibnamefont {Sendouda}}, \ and\
  \bibinfo {author} {\bibfnamefont {J.}~\bibnamefont {Yokoyama}},\ }\href@noop
  {} {\  (\bibinfo {year} {2020})},\ \Eprint {http://arxiv.org/abs/2002.12778}
  {arXiv:2002.12778 [astro-ph.CO]} \BibitemShut {NoStop}%
\bibitem [{\citenamefont {Chiba}\ and\ \citenamefont
  {Yokoyama}(2017)}]{Chiba:2017rvs}%
  \BibitemOpen
  \bibfield  {author} {\bibinfo {author} {\bibfnamefont {T.}~\bibnamefont
  {Chiba}}\ and\ \bibinfo {author} {\bibfnamefont {S.}~\bibnamefont
  {Yokoyama}},\ }\href {\doibase 10.1093/ptep/ptx087} {\bibfield  {journal}
  {\bibinfo  {journal} {PTEP}\ }\textbf {\bibinfo {volume} {2017}},\ \bibinfo
  {pages} {083E01} (\bibinfo {year} {2017})},\ \Eprint
  {http://arxiv.org/abs/1704.06573} {arXiv:1704.06573 [gr-qc]} \BibitemShut
  {NoStop}%
%%CITATION = ARXIV:1704.06573;%%
\bibitem [{\citenamefont {De~Luca}\ \emph {et~al.}(2019)\citenamefont
  {De~Luca}, \citenamefont {Desjacques}, \citenamefont {Franciolini},
  \citenamefont {Malhotra},\ and\ \citenamefont {Riotto}}]{DeLuca:2019buf}%
  \BibitemOpen
  \bibfield  {author} {\bibinfo {author} {\bibfnamefont {V.}~\bibnamefont
  {De~Luca}}, \bibinfo {author} {\bibfnamefont {V.}~\bibnamefont {Desjacques}},
  \bibinfo {author} {\bibfnamefont {G.}~\bibnamefont {Franciolini}}, \bibinfo
  {author} {\bibfnamefont {A.}~\bibnamefont {Malhotra}}, \ and\ \bibinfo
  {author} {\bibfnamefont {A.}~\bibnamefont {Riotto}},\ }\href {\doibase
  10.1088/1475-7516/2019/05/018} {\bibfield  {journal} {\bibinfo  {journal}
  {JCAP}\ }\textbf {\bibinfo {volume} {05}},\ \bibinfo {pages} {018} (\bibinfo
  {year} {2019})},\ \Eprint {http://arxiv.org/abs/1903.01179} {arXiv:1903.01179
  [astro-ph.CO]} \BibitemShut {NoStop}%
\bibitem [{\citenamefont {Mirbabayi}\ \emph {et~al.}(2020)\citenamefont
  {Mirbabayi}, \citenamefont {Gruzinov},\ and\ \citenamefont
  {Nore\~na}}]{Mirbabayi:2019uph}%
  \BibitemOpen
  \bibfield  {author} {\bibinfo {author} {\bibfnamefont {M.}~\bibnamefont
  {Mirbabayi}}, \bibinfo {author} {\bibfnamefont {A.}~\bibnamefont {Gruzinov}},
  \ and\ \bibinfo {author} {\bibfnamefont {J.}~\bibnamefont {Nore\~na}},\
  }\href {\doibase 10.1088/1475-7516/2020/03/017} {\bibfield  {journal}
  {\bibinfo  {journal} {JCAP}\ }\textbf {\bibinfo {volume} {03}},\ \bibinfo
  {pages} {017} (\bibinfo {year} {2020})},\ \Eprint
  {http://arxiv.org/abs/1901.05963} {arXiv:1901.05963 [astro-ph.CO]}
  \BibitemShut {NoStop}%
\bibitem [{\citenamefont {Amendola}\ \emph {et~al.}(2018)\citenamefont
  {Amendola}, \citenamefont {Rubio},\ and\ \citenamefont
  {Wetterich}}]{Amendola:2017xhl}%
  \BibitemOpen
  \bibfield  {author} {\bibinfo {author} {\bibfnamefont {L.}~\bibnamefont
  {Amendola}}, \bibinfo {author} {\bibfnamefont {J.}~\bibnamefont {Rubio}}, \
  and\ \bibinfo {author} {\bibfnamefont {C.}~\bibnamefont {Wetterich}},\ }\href
  {\doibase 10.1103/PhysRevD.97.081302} {\bibfield  {journal} {\bibinfo
  {journal} {Phys. Rev. D}\ }\textbf {\bibinfo {volume} {97}},\ \bibinfo
  {pages} {081302} (\bibinfo {year} {2018})},\ \Eprint
  {http://arxiv.org/abs/1711.09915} {arXiv:1711.09915 [astro-ph.CO]}
  \BibitemShut {NoStop}%
\bibitem [{\citenamefont {Dom\`enech}\ and\ \citenamefont
  {Sasaki}(2021)}]{Domenech:2021uyx}%
  \BibitemOpen
  \bibfield  {author} {\bibinfo {author} {\bibfnamefont {G.}~\bibnamefont
  {Dom\`enech}}\ and\ \bibinfo {author} {\bibfnamefont {M.}~\bibnamefont
  {Sasaki}},\ }\href@noop {} {\  (\bibinfo {year} {2021})},\ \Eprint
  {http://arxiv.org/abs/2104.05271} {arXiv:2104.05271 [hep-th]} \BibitemShut
  {NoStop}%
\bibitem [{\citenamefont {Harada}\ \emph {et~al.}(2016)\citenamefont {Harada},
  \citenamefont {Yoo}, \citenamefont {Kohri}, \citenamefont {Nakao},\ and\
  \citenamefont {Jhingan}}]{Harada:2016mhb}%
  \BibitemOpen
  \bibfield  {author} {\bibinfo {author} {\bibfnamefont {T.}~\bibnamefont
  {Harada}}, \bibinfo {author} {\bibfnamefont {C.-M.}\ \bibnamefont {Yoo}},
  \bibinfo {author} {\bibfnamefont {K.}~\bibnamefont {Kohri}}, \bibinfo
  {author} {\bibfnamefont {K.-i.}\ \bibnamefont {Nakao}}, \ and\ \bibinfo
  {author} {\bibfnamefont {S.}~\bibnamefont {Jhingan}},\ }\href {\doibase
  10.3847/1538-4357/833/1/61} {\bibfield  {journal} {\bibinfo  {journal}
  {Astrophys. J.}\ }\textbf {\bibinfo {volume} {833}},\ \bibinfo {pages} {61}
  (\bibinfo {year} {2016})},\ \Eprint {http://arxiv.org/abs/1609.01588}
  {arXiv:1609.01588 [astro-ph.CO]} \BibitemShut {NoStop}%
%%CITATION = ARXIV:1609.01588;%%
\bibitem [{\citenamefont {Kokubu}\ \emph {et~al.}(2018)\citenamefont {Kokubu},
  \citenamefont {Kyutoku}, \citenamefont {Kohri},\ and\ \citenamefont
  {Harada}}]{Kokubu:2018fxy}%
  \BibitemOpen
  \bibfield  {author} {\bibinfo {author} {\bibfnamefont {T.}~\bibnamefont
  {Kokubu}}, \bibinfo {author} {\bibfnamefont {K.}~\bibnamefont {Kyutoku}},
  \bibinfo {author} {\bibfnamefont {K.}~\bibnamefont {Kohri}}, \ and\ \bibinfo
  {author} {\bibfnamefont {T.}~\bibnamefont {Harada}},\ }\href {\doibase
  10.1103/PhysRevD.98.123024} {\bibfield  {journal} {\bibinfo  {journal} {Phys.
  Rev. D}\ }\textbf {\bibinfo {volume} {98}},\ \bibinfo {pages} {123024}
  (\bibinfo {year} {2018})},\ \Eprint {http://arxiv.org/abs/1810.03490}
  {arXiv:1810.03490 [astro-ph.CO]} \BibitemShut {NoStop}%
\bibitem [{\citenamefont {De~Luca}\ \emph {et~al.}(2020)\citenamefont
  {De~Luca}, \citenamefont {Franciolini}, \citenamefont {Pani},\ and\
  \citenamefont {Riotto}}]{DeLuca:2020bjf}%
  \BibitemOpen
  \bibfield  {author} {\bibinfo {author} {\bibfnamefont {V.}~\bibnamefont
  {De~Luca}}, \bibinfo {author} {\bibfnamefont {G.}~\bibnamefont
  {Franciolini}}, \bibinfo {author} {\bibfnamefont {P.}~\bibnamefont {Pani}}, \
  and\ \bibinfo {author} {\bibfnamefont {A.}~\bibnamefont {Riotto}},\ }\href
  {\doibase 10.1088/1475-7516/2020/04/052} {\bibfield  {journal} {\bibinfo
  {journal} {JCAP}\ }\textbf {\bibinfo {volume} {04}},\ \bibinfo {pages} {052}
  (\bibinfo {year} {2020})},\ \Eprint {http://arxiv.org/abs/2003.02778}
  {arXiv:2003.02778 [astro-ph.CO]} \BibitemShut {NoStop}%
\bibitem [{\citenamefont {Fishbach}\ \emph {et~al.}(2017)\citenamefont
  {Fishbach}, \citenamefont {Holz},\ and\ \citenamefont
  {Farr}}]{Fishbach:2017dwv}%
  \BibitemOpen
  \bibfield  {author} {\bibinfo {author} {\bibfnamefont {M.}~\bibnamefont
  {Fishbach}}, \bibinfo {author} {\bibfnamefont {D.~E.}\ \bibnamefont {Holz}},
  \ and\ \bibinfo {author} {\bibfnamefont {B.}~\bibnamefont {Farr}},\ }\href
  {\doibase 10.3847/2041-8213/aa7045} {\bibfield  {journal} {\bibinfo
  {journal} {Astrophys. J. Lett.}\ }\textbf {\bibinfo {volume} {840}},\
  \bibinfo {pages} {L24} (\bibinfo {year} {2017})},\ \Eprint
  {http://arxiv.org/abs/1703.06869} {arXiv:1703.06869 [astro-ph.HE]}
  \BibitemShut {NoStop}%
\bibitem [{\citenamefont {Arbey}\ \emph
  {et~al.}(2020{\natexlab{a}})\citenamefont {Arbey}, \citenamefont
  {Auffinger},\ and\ \citenamefont {Silk}}]{Arbey:2019vqx}%
  \BibitemOpen
  \bibfield  {author} {\bibinfo {author} {\bibfnamefont {A.}~\bibnamefont
  {Arbey}}, \bibinfo {author} {\bibfnamefont {J.}~\bibnamefont {Auffinger}}, \
  and\ \bibinfo {author} {\bibfnamefont {J.}~\bibnamefont {Silk}},\ }\href
  {\doibase 10.1103/PhysRevD.101.023010} {\bibfield  {journal} {\bibinfo
  {journal} {Phys. Rev. D}\ }\textbf {\bibinfo {volume} {101}},\ \bibinfo
  {pages} {023010} (\bibinfo {year} {2020}{\natexlab{a}})},\ \Eprint
  {http://arxiv.org/abs/1906.04750} {arXiv:1906.04750 [astro-ph.CO]}
  \BibitemShut {NoStop}%
\bibitem [{\citenamefont {Dong}\ \emph {et~al.}(2016)\citenamefont {Dong},
  \citenamefont {Kinney},\ and\ \citenamefont {Stojkovic}}]{Dong:2015yjs}%
  \BibitemOpen
  \bibfield  {author} {\bibinfo {author} {\bibfnamefont {R.}~\bibnamefont
  {Dong}}, \bibinfo {author} {\bibfnamefont {W.~H.}\ \bibnamefont {Kinney}}, \
  and\ \bibinfo {author} {\bibfnamefont {D.}~\bibnamefont {Stojkovic}},\ }\href
  {\doibase 10.1088/1475-7516/2016/10/034} {\bibfield  {journal} {\bibinfo
  {journal} {JCAP}\ }\textbf {\bibinfo {volume} {10}},\ \bibinfo {pages} {034}
  (\bibinfo {year} {2016})},\ \Eprint {http://arxiv.org/abs/1511.05642}
  {arXiv:1511.05642 [astro-ph.CO]} \BibitemShut {NoStop}%
\bibitem [{\citenamefont {Kuhnel}(2020)}]{Kuhnel:2019zbc}%
  \BibitemOpen
  \bibfield  {author} {\bibinfo {author} {\bibfnamefont {F.}~\bibnamefont
  {Kuhnel}},\ }\href {\doibase 10.1140/epjc/s10052-020-7807-z} {\bibfield
  {journal} {\bibinfo  {journal} {Eur. Phys. J. C}\ }\textbf {\bibinfo {volume}
  {80}},\ \bibinfo {pages} {243} (\bibinfo {year} {2020})},\ \Eprint
  {http://arxiv.org/abs/1909.04742} {arXiv:1909.04742 [astro-ph.CO]}
  \BibitemShut {NoStop}%
\bibitem [{\citenamefont {Arbey}\ \emph
  {et~al.}(2020{\natexlab{b}})\citenamefont {Arbey}, \citenamefont
  {Auffinger},\ and\ \citenamefont {Silk}}]{Arbey:2019jmj}%
  \BibitemOpen
  \bibfield  {author} {\bibinfo {author} {\bibfnamefont {A.}~\bibnamefont
  {Arbey}}, \bibinfo {author} {\bibfnamefont {J.}~\bibnamefont {Auffinger}}, \
  and\ \bibinfo {author} {\bibfnamefont {J.}~\bibnamefont {Silk}},\ }\href
  {\doibase 10.1093/mnras/staa765} {\bibfield  {journal} {\bibinfo  {journal}
  {Mon. Not. Roy. Astron. Soc.}\ }\textbf {\bibinfo {volume} {494}},\ \bibinfo
  {pages} {1257} (\bibinfo {year} {2020}{\natexlab{b}})},\ \Eprint
  {http://arxiv.org/abs/1906.04196} {arXiv:1906.04196 [astro-ph.CO]}
  \BibitemShut {NoStop}%
\bibitem [{\citenamefont {Bai}\ and\ \citenamefont
  {Orlofsky}(2020)}]{Bai:2019zcd}%
  \BibitemOpen
  \bibfield  {author} {\bibinfo {author} {\bibfnamefont {Y.}~\bibnamefont
  {Bai}}\ and\ \bibinfo {author} {\bibfnamefont {N.}~\bibnamefont {Orlofsky}},\
  }\href {\doibase 10.1103/PhysRevD.101.055006} {\bibfield  {journal} {\bibinfo
   {journal} {Phys. Rev. D}\ }\textbf {\bibinfo {volume} {101}},\ \bibinfo
  {pages} {055006} (\bibinfo {year} {2020})},\ \Eprint
  {http://arxiv.org/abs/1906.04858} {arXiv:1906.04858 [hep-ph]} \BibitemShut
  {NoStop}%
\bibitem [{\citenamefont {Dasgupta}\ \emph {et~al.}(2020)\citenamefont
  {Dasgupta}, \citenamefont {Laha},\ and\ \citenamefont
  {Ray}}]{Dasgupta:2019cae}%
  \BibitemOpen
  \bibfield  {author} {\bibinfo {author} {\bibfnamefont {B.}~\bibnamefont
  {Dasgupta}}, \bibinfo {author} {\bibfnamefont {R.}~\bibnamefont {Laha}}, \
  and\ \bibinfo {author} {\bibfnamefont {A.}~\bibnamefont {Ray}},\ }\href
  {\doibase 10.1103/PhysRevLett.125.101101} {\bibfield  {journal} {\bibinfo
  {journal} {Phys. Rev. Lett.}\ }\textbf {\bibinfo {volume} {125}},\ \bibinfo
  {pages} {101101} (\bibinfo {year} {2020})},\ \Eprint
  {http://arxiv.org/abs/1912.01014} {arXiv:1912.01014 [hep-ph]} \BibitemShut
  {NoStop}%
\bibitem [{\citenamefont {Page}(1976{\natexlab{b}})}]{Page:1976ki}%
  \BibitemOpen
  \bibfield  {author} {\bibinfo {author} {\bibfnamefont {D.~N.}\ \bibnamefont
  {Page}},\ }\href {\doibase 10.1103/PhysRevD.14.3260} {\bibfield  {journal}
  {\bibinfo  {journal} {Phys. Rev. D}\ }\textbf {\bibinfo {volume} {14}},\
  \bibinfo {pages} {3260} (\bibinfo {year} {1976}{\natexlab{b}})}\BibitemShut
  {NoStop}%
\bibitem [{\citenamefont {Taylor}\ \emph {et~al.}(1998)\citenamefont {Taylor},
  \citenamefont {Chambers},\ and\ \citenamefont {Hiscock}}]{Taylor:1998dk}%
  \BibitemOpen
  \bibfield  {author} {\bibinfo {author} {\bibfnamefont {B.~E.}\ \bibnamefont
  {Taylor}}, \bibinfo {author} {\bibfnamefont {C.~M.}\ \bibnamefont
  {Chambers}}, \ and\ \bibinfo {author} {\bibfnamefont {W.~A.}\ \bibnamefont
  {Hiscock}},\ }\href {\doibase 10.1103/PhysRevD.58.044012} {\bibfield
  {journal} {\bibinfo  {journal} {Phys. Rev. D}\ }\textbf {\bibinfo {volume}
  {58}},\ \bibinfo {pages} {044012} (\bibinfo {year} {1998})},\ \Eprint
  {http://arxiv.org/abs/gr-qc/9801044} {arXiv:gr-qc/9801044} \BibitemShut
  {NoStop}%
\bibitem [{\citenamefont {Arbey}\ \emph {et~al.}(2021)\citenamefont {Arbey},
  \citenamefont {Auffinger}, \citenamefont {Sandick}, \citenamefont {Shams
  Es~Haghi},\ and\ \citenamefont {Sinha}}]{Arbey:2021ysg}%
  \BibitemOpen
  \bibfield  {author} {\bibinfo {author} {\bibfnamefont {A.}~\bibnamefont
  {Arbey}}, \bibinfo {author} {\bibfnamefont {J.}~\bibnamefont {Auffinger}},
  \bibinfo {author} {\bibfnamefont {P.}~\bibnamefont {Sandick}}, \bibinfo
  {author} {\bibfnamefont {B.}~\bibnamefont {Shams Es~Haghi}}, \ and\ \bibinfo
  {author} {\bibfnamefont {K.}~\bibnamefont {Sinha}},\ }\href@noop {} {\
  (\bibinfo {year} {2021})},\ \Eprint {http://arxiv.org/abs/2104.04051}
  {arXiv:2104.04051 [astro-ph.CO]} \BibitemShut {NoStop}%
\bibitem [{\citenamefont {Masina}(2021)}]{Masina:2021zpu}%
  \BibitemOpen
  \bibfield  {author} {\bibinfo {author} {\bibfnamefont {I.}~\bibnamefont
  {Masina}},\ }\href@noop {} {\  (\bibinfo {year} {2021})},\ \Eprint
  {http://arxiv.org/abs/2103.13825} {arXiv:2103.13825 [gr-qc]} \BibitemShut
  {NoStop}%
\bibitem [{\citenamefont {Bisnovatyi-Kogan}\ and\ \citenamefont
  {Rudenko}(2004)}]{BisnovatyiKogan:2004bk}%
  \BibitemOpen
  \bibfield  {author} {\bibinfo {author} {\bibfnamefont {G.~S.}\ \bibnamefont
  {Bisnovatyi-Kogan}}\ and\ \bibinfo {author} {\bibfnamefont {V.~N.}\
  \bibnamefont {Rudenko}},\ }\href {\doibase 10.1088/0264-9381/21/14/001}
  {\bibfield  {journal} {\bibinfo  {journal} {Class. Quant. Grav.}\ }\textbf
  {\bibinfo {volume} {21}},\ \bibinfo {pages} {3347} (\bibinfo {year}
  {2004})},\ \Eprint {http://arxiv.org/abs/gr-qc/0406089} {arXiv:gr-qc/0406089}
  \BibitemShut {NoStop}%
\bibitem [{\citenamefont {Papanikolaou}\ \emph {et~al.}(2021)\citenamefont
  {Papanikolaou}, \citenamefont {Vennin},\ and\ \citenamefont
  {Langlois}}]{Papanikolaou:2020qtd}%
  \BibitemOpen
  \bibfield  {author} {\bibinfo {author} {\bibfnamefont {T.}~\bibnamefont
  {Papanikolaou}}, \bibinfo {author} {\bibfnamefont {V.}~\bibnamefont
  {Vennin}}, \ and\ \bibinfo {author} {\bibfnamefont {D.}~\bibnamefont
  {Langlois}},\ }\href {\doibase 10.1088/1475-7516/2021/03/053} {\bibfield
  {journal} {\bibinfo  {journal} {JCAP}\ }\textbf {\bibinfo {volume} {03}},\
  \bibinfo {pages} {053} (\bibinfo {year} {2021})},\ \Eprint
  {http://arxiv.org/abs/2010.11573} {arXiv:2010.11573 [astro-ph.CO]}
  \BibitemShut {NoStop}%
\bibitem [{\citenamefont {Inomata}\ \emph
  {et~al.}(2019{\natexlab{a}})\citenamefont {Inomata}, \citenamefont {Kohri},
  \citenamefont {Nakama},\ and\ \citenamefont {Terada}}]{Inomata:2019ivs}%
  \BibitemOpen
  \bibfield  {author} {\bibinfo {author} {\bibfnamefont {K.}~\bibnamefont
  {Inomata}}, \bibinfo {author} {\bibfnamefont {K.}~\bibnamefont {Kohri}},
  \bibinfo {author} {\bibfnamefont {T.}~\bibnamefont {Nakama}}, \ and\ \bibinfo
  {author} {\bibfnamefont {T.}~\bibnamefont {Terada}},\ }\href {\doibase
  10.1103/PhysRevD.100.043532} {\bibfield  {journal} {\bibinfo  {journal}
  {Phys. Rev. D}\ }\textbf {\bibinfo {volume} {100}},\ \bibinfo {pages}
  {043532} (\bibinfo {year} {2019}{\natexlab{a}})},\ \Eprint
  {http://arxiv.org/abs/1904.12879} {arXiv:1904.12879 [astro-ph.CO]}
  \BibitemShut {NoStop}%
\bibitem [{\citenamefont {Dom\`enech}\ \emph {et~al.}(2021)\citenamefont
  {Dom\`enech}, \citenamefont {Lin},\ and\ \citenamefont
  {Sasaki}}]{Domenech:2020ssp}%
  \BibitemOpen
  \bibfield  {author} {\bibinfo {author} {\bibfnamefont {G.}~\bibnamefont
  {Dom\`enech}}, \bibinfo {author} {\bibfnamefont {C.}~\bibnamefont {Lin}}, \
  and\ \bibinfo {author} {\bibfnamefont {M.}~\bibnamefont {Sasaki}},\ }\href
  {\doibase 10.1088/1475-7516/2021/04/062} {\bibfield  {journal} {\bibinfo
  {journal} {JCAP}\ }\textbf {\bibinfo {volume} {04}},\ \bibinfo {pages} {062}
  (\bibinfo {year} {2021})},\ \Eprint {http://arxiv.org/abs/2012.08151}
  {arXiv:2012.08151 [gr-qc]} \BibitemShut {NoStop}%
\bibitem [{\citenamefont {Inomata}\ \emph {et~al.}(2020)\citenamefont
  {Inomata}, \citenamefont {Kawasaki}, \citenamefont {Mukaida}, \citenamefont
  {Terada},\ and\ \citenamefont {Yanagida}}]{Inomata:2020lmk}%
  \BibitemOpen
  \bibfield  {author} {\bibinfo {author} {\bibfnamefont {K.}~\bibnamefont
  {Inomata}}, \bibinfo {author} {\bibfnamefont {M.}~\bibnamefont {Kawasaki}},
  \bibinfo {author} {\bibfnamefont {K.}~\bibnamefont {Mukaida}}, \bibinfo
  {author} {\bibfnamefont {T.}~\bibnamefont {Terada}}, \ and\ \bibinfo {author}
  {\bibfnamefont {T.~T.}\ \bibnamefont {Yanagida}},\ }\href {\doibase
  10.1103/PhysRevD.101.123533} {\bibfield  {journal} {\bibinfo  {journal}
  {Phys. Rev. D}\ }\textbf {\bibinfo {volume} {101}},\ \bibinfo {pages}
  {123533} (\bibinfo {year} {2020})},\ \Eprint
  {http://arxiv.org/abs/2003.10455} {arXiv:2003.10455 [astro-ph.CO]}
  \BibitemShut {NoStop}%
\bibitem [{\citenamefont {Aghanim}\ \emph {et~al.}(2018)\citenamefont {Aghanim}
  \emph {et~al.}}]{Aghanim:2018eyx}%
  \BibitemOpen
  \bibfield  {author} {\bibinfo {author} {\bibfnamefont {N.}~\bibnamefont
  {Aghanim}} \emph {et~al.} (\bibinfo {collaboration} {Planck}),\ }\href@noop
  {} {\  (\bibinfo {year} {2018})},\ \Eprint {http://arxiv.org/abs/1807.06209}
  {arXiv:1807.06209 [astro-ph.CO]} \BibitemShut {NoStop}%
%%CITATION = ARXIV:1807.06209;%%
\bibitem [{\citenamefont {Abazajian}\ \emph {et~al.}(2016)\citenamefont
  {Abazajian} \emph {et~al.}}]{Abazajian:2016yjj}%
  \BibitemOpen
  \bibfield  {author} {\bibinfo {author} {\bibfnamefont {K.~N.}\ \bibnamefont
  {Abazajian}} \emph {et~al.} (\bibinfo {collaboration} {CMB-S4}),\ }\href@noop
  {} {\  (\bibinfo {year} {2016})},\ \Eprint {http://arxiv.org/abs/1610.02743}
  {arXiv:1610.02743 [astro-ph.CO]} \BibitemShut {NoStop}%
\bibitem [{\citenamefont {Thrane}\ and\ \citenamefont
  {Romano}(2013)}]{Thrane:2013oya}%
  \BibitemOpen
  \bibfield  {author} {\bibinfo {author} {\bibfnamefont {E.}~\bibnamefont
  {Thrane}}\ and\ \bibinfo {author} {\bibfnamefont {J.~D.}\ \bibnamefont
  {Romano}},\ }\href {\doibase 10.1103/PhysRevD.88.124032} {\bibfield
  {journal} {\bibinfo  {journal} {Phys. Rev. D}\ }\textbf {\bibinfo {volume}
  {88}},\ \bibinfo {pages} {124032} (\bibinfo {year} {2013})},\ \Eprint
  {http://arxiv.org/abs/1310.5300} {arXiv:1310.5300 [astro-ph.IM]} \BibitemShut
  {NoStop}%
\bibitem [{\citenamefont {Abbott}\ \emph {et~al.}(2021)\citenamefont {Abbott}
  \emph {et~al.}}]{Abbott:2021xxi}%
  \BibitemOpen
  \bibfield  {author} {\bibinfo {author} {\bibfnamefont {R.}~\bibnamefont
  {Abbott}} \emph {et~al.} (\bibinfo {collaboration} {LIGO Scientific, Virgo,
  KAGRA}),\ }\href@noop {} {\  (\bibinfo {year} {2021})},\ \Eprint
  {http://arxiv.org/abs/2101.12130} {arXiv:2101.12130 [gr-qc]} \BibitemShut
  {NoStop}%
\bibitem [{\citenamefont {Tomita}(1967)}]{tomita}%
  \BibitemOpen
  \bibfield  {author} {\bibinfo {author} {\bibfnamefont {K.}~\bibnamefont
  {Tomita}},\ }\href {\doibase 10.1143/PTP.37.831} {\bibfield  {journal}
  {\bibinfo  {journal} {Progress of Theoretical Physics}\ }\textbf {\bibinfo
  {volume} {37}},\ \bibinfo {pages} {831} (\bibinfo {year} {1967})},\ \Eprint
  {http://arxiv.org/abs/https://academic.oup.com/ptp/article-pdf/37/5/831/5234391/37-5-831.pdf}
  {https://academic.oup.com/ptp/article-pdf/37/5/831/5234391/37-5-831.pdf}
  \BibitemShut {NoStop}%
\bibitem [{\citenamefont {Matarrese}\ \emph {et~al.}(1993)\citenamefont
  {Matarrese}, \citenamefont {Pantano},\ and\ \citenamefont
  {Saez}}]{Matarrese:1992rp}%
  \BibitemOpen
  \bibfield  {author} {\bibinfo {author} {\bibfnamefont {S.}~\bibnamefont
  {Matarrese}}, \bibinfo {author} {\bibfnamefont {O.}~\bibnamefont {Pantano}},
  \ and\ \bibinfo {author} {\bibfnamefont {D.}~\bibnamefont {Saez}},\ }\href
  {\doibase 10.1103/PhysRevD.47.1311} {\bibfield  {journal} {\bibinfo
  {journal} {Phys. Rev. D}\ }\textbf {\bibinfo {volume} {47}},\ \bibinfo
  {pages} {1311} (\bibinfo {year} {1993})}\BibitemShut {NoStop}%
\bibitem [{\citenamefont {Matarrese}\ \emph {et~al.}(1994)\citenamefont
  {Matarrese}, \citenamefont {Pantano},\ and\ \citenamefont
  {Saez}}]{Matarrese:1993zf}%
  \BibitemOpen
  \bibfield  {author} {\bibinfo {author} {\bibfnamefont {S.}~\bibnamefont
  {Matarrese}}, \bibinfo {author} {\bibfnamefont {O.}~\bibnamefont {Pantano}},
  \ and\ \bibinfo {author} {\bibfnamefont {D.}~\bibnamefont {Saez}},\ }\href
  {\doibase 10.1103/PhysRevLett.72.320} {\bibfield  {journal} {\bibinfo
  {journal} {Phys. Rev. Lett.}\ }\textbf {\bibinfo {volume} {72}},\ \bibinfo
  {pages} {320} (\bibinfo {year} {1994})},\ \Eprint
  {http://arxiv.org/abs/astro-ph/9310036} {arXiv:astro-ph/9310036} \BibitemShut
  {NoStop}%
\bibitem [{\citenamefont {Ananda}\ \emph {et~al.}(2007)\citenamefont {Ananda},
  \citenamefont {Clarkson},\ and\ \citenamefont {Wands}}]{Ananda:2006af}%
  \BibitemOpen
  \bibfield  {author} {\bibinfo {author} {\bibfnamefont {K.~N.}\ \bibnamefont
  {Ananda}}, \bibinfo {author} {\bibfnamefont {C.}~\bibnamefont {Clarkson}}, \
  and\ \bibinfo {author} {\bibfnamefont {D.}~\bibnamefont {Wands}},\ }\href
  {\doibase 10.1103/PhysRevD.75.123518} {\bibfield  {journal} {\bibinfo
  {journal} {Phys. Rev. D}\ }\textbf {\bibinfo {volume} {75}},\ \bibinfo
  {pages} {123518} (\bibinfo {year} {2007})},\ \Eprint
  {http://arxiv.org/abs/gr-qc/0612013} {arXiv:gr-qc/0612013} \BibitemShut
  {NoStop}%
\bibitem [{\citenamefont {Baumann}\ \emph {et~al.}(2007)\citenamefont
  {Baumann}, \citenamefont {Steinhardt}, \citenamefont {Takahashi},\ and\
  \citenamefont {Ichiki}}]{Baumann:2007zm}%
  \BibitemOpen
  \bibfield  {author} {\bibinfo {author} {\bibfnamefont {D.}~\bibnamefont
  {Baumann}}, \bibinfo {author} {\bibfnamefont {P.~J.}\ \bibnamefont
  {Steinhardt}}, \bibinfo {author} {\bibfnamefont {K.}~\bibnamefont
  {Takahashi}}, \ and\ \bibinfo {author} {\bibfnamefont {K.}~\bibnamefont
  {Ichiki}},\ }\href {\doibase 10.1103/PhysRevD.76.084019} {\bibfield
  {journal} {\bibinfo  {journal} {Phys. Rev. D}\ }\textbf {\bibinfo {volume}
  {76}},\ \bibinfo {pages} {084019} (\bibinfo {year} {2007})},\ \Eprint
  {http://arxiv.org/abs/hep-th/0703290} {arXiv:hep-th/0703290} \BibitemShut
  {NoStop}%
\bibitem [{\citenamefont {Saito}\ and\ \citenamefont
  {Yokoyama}(2010)}]{Saito:2009jt}%
  \BibitemOpen
  \bibfield  {author} {\bibinfo {author} {\bibfnamefont {R.}~\bibnamefont
  {Saito}}\ and\ \bibinfo {author} {\bibfnamefont {J.}~\bibnamefont
  {Yokoyama}},\ }\href {\doibase 10.1143/PTP.126.351} {\bibfield  {journal}
  {\bibinfo  {journal} {Prog. Theor. Phys.}\ }\textbf {\bibinfo {volume}
  {123}},\ \bibinfo {pages} {867} (\bibinfo {year} {2010})},\ \bibinfo {note}
  {[Erratum: Prog.Theor.Phys. 126, 351--352 (2011)]},\ \Eprint
  {http://arxiv.org/abs/0912.5317} {arXiv:0912.5317 [astro-ph.CO]} \BibitemShut
  {NoStop}%
\bibitem [{\citenamefont {Kodama}\ and\ \citenamefont
  {Sasaki}(1986)}]{Kodama:1986fg}%
  \BibitemOpen
  \bibfield  {author} {\bibinfo {author} {\bibfnamefont {H.}~\bibnamefont
  {Kodama}}\ and\ \bibinfo {author} {\bibfnamefont {M.}~\bibnamefont
  {Sasaki}},\ }\href {\doibase 10.1142/S0217751X86000137} {\bibfield  {journal}
  {\bibinfo  {journal} {Int. J. Mod. Phys. A}\ }\textbf {\bibinfo {volume}
  {1}},\ \bibinfo {pages} {265} (\bibinfo {year} {1986})}\BibitemShut {NoStop}%
\bibitem [{\citenamefont {Kodama}\ and\ \citenamefont
  {Sasaki}(1987)}]{Kodama:1986ud}%
  \BibitemOpen
  \bibfield  {author} {\bibinfo {author} {\bibfnamefont {H.}~\bibnamefont
  {Kodama}}\ and\ \bibinfo {author} {\bibfnamefont {M.}~\bibnamefont
  {Sasaki}},\ }\href {\doibase 10.1142/S0217751X8700020X} {\bibfield  {journal}
  {\bibinfo  {journal} {Int. J. Mod. Phys. A}\ }\textbf {\bibinfo {volume}
  {2}},\ \bibinfo {pages} {491} (\bibinfo {year} {1987})}\BibitemShut {NoStop}%
\bibitem [{\citenamefont {Caprini}\ and\ \citenamefont
  {Figueroa}(2018)}]{Caprini:2018mtu}%
  \BibitemOpen
  \bibfield  {author} {\bibinfo {author} {\bibfnamefont {C.}~\bibnamefont
  {Caprini}}\ and\ \bibinfo {author} {\bibfnamefont {D.~G.}\ \bibnamefont
  {Figueroa}},\ }\href {\doibase 10.1088/1361-6382/aac608} {\bibfield
  {journal} {\bibinfo  {journal} {Class. Quant. Grav.}\ }\textbf {\bibinfo
  {volume} {35}},\ \bibinfo {pages} {163001} (\bibinfo {year} {2018})},\
  \Eprint {http://arxiv.org/abs/1801.04268} {arXiv:1801.04268 [astro-ph.CO]}
  \BibitemShut {NoStop}%
\bibitem [{\citenamefont {Inomata}\ \emph
  {et~al.}(2019{\natexlab{b}})\citenamefont {Inomata}, \citenamefont {Kohri},
  \citenamefont {Nakama},\ and\ \citenamefont {Terada}}]{Inomata:2019zqy}%
  \BibitemOpen
  \bibfield  {author} {\bibinfo {author} {\bibfnamefont {K.}~\bibnamefont
  {Inomata}}, \bibinfo {author} {\bibfnamefont {K.}~\bibnamefont {Kohri}},
  \bibinfo {author} {\bibfnamefont {T.}~\bibnamefont {Nakama}}, \ and\ \bibinfo
  {author} {\bibfnamefont {T.}~\bibnamefont {Terada}},\ }\href {\doibase
  10.1088/1475-7516/2019/10/071} {\bibfield  {journal} {\bibinfo  {journal}
  {JCAP}\ }\textbf {\bibinfo {volume} {10}},\ \bibinfo {pages} {071} (\bibinfo
  {year} {2019}{\natexlab{b}})},\ \Eprint {http://arxiv.org/abs/1904.12878}
  {arXiv:1904.12878 [astro-ph.CO]} \BibitemShut {NoStop}%
\end{thebibliography}%

\end{document}